\documentclass[prb,superscriptaddress,twocolumn,showpacs,preprintnumbers,amsmath,floatfix]{revtex4}
\usepackage{amssymb}
\usepackage{graphicx}
\usepackage{color}

\begin{document}

\title{Distinct Nature of Orbital Selective Mott Phases Dominated by the Low-energy Local Spin Fluctuations}
\author{Ze-Yi Song}
\affiliation{Shanghai Key Laboratory of Special Artificial Microstructure Materials and Technology, School of Physics Science and engineering, Tongji University, Shanghai 200092, P.R. China}
\author{Xiu-Cai Jiang}
\affiliation{Shanghai Key Laboratory of Special Artificial Microstructure Materials and Technology, School of Physics Science and engineering, Tongji University, Shanghai 200092, P.R. China}
\author{Hai-Qing Lin}
\affiliation{Beijing Computational Science Research Center, Beijing 100084, P.R. China}
\author{Yu-Zhong Zhang}
\email[Corresponding author.]{Email: yzzhang@tongji.edu.cn}
\affiliation{Shanghai Key Laboratory of Special Artificial Microstructure Materials and Technology, School of Physics Science and engineering, Tongji University, Shanghai 200092, P.R. China}
\affiliation{Beijing Computational Science Research Center, Beijing 100084, P.R. China}
\date{\today}

\begin{abstract}
\centerline{Abstract}
Quantum orbital selective Mott (OSM) transitions are investigated within dynamical mean-field theory based on a two-orbital Hubbard model with different bandwidth at half filling. We find two distinct OSM phases both showing coexistence of itinerant electrons and localized spins, dependent on whether the Hund's coupling is full or of Ising type. The critical values and the nature of the OSM transitions are efficiently determined by entanglement entropy. We reveal that vanishing of the Kondo energy scale evidenced by absence of local spin fluctuations at low frequency in local dynamical spin susceptibility is responsible for the appearance of non-Fermi-liquid OSM phase in Ising Hund's coupling case. We argue that this scenario can also be applied to account for emergent quantum non-Fermi liquid in one-band Hubbard model when short-range antiferromagnetic order is considered.
\end{abstract}

\maketitle

\section{ Introduction }

Mott metal-insulator transition in a single-band Hubbard model at half-filling, realized in parent compound of high-T$_c$ cuprates like La$_2$CuO$_4$~\cite{DagottoRMP1994,ImadaRMP1998}, has been extensively studied and a clear physics picture has emerged with the development of dynamical mean-field theory (DMFT) which is exact in infinite dimensions~\cite{ImadaRMP1998,GeorgesRMP1996}. However, many correlated materials with more than one orbital across the Fermi level can not be described by the one-band Hubbard model. It has been shown that the orbital degrees of freedom together with the Hund's coupling, which are two important ingredients in the multi-band Hubbard model, may be responsible for many anomalous properties observed in topical multi-orbital materials like iron-based superconductors~\cite{HauleNJP2009,YinNP2011}, ruthenates~\cite{WernerPRL2008,MravljePRL2011} and other 4d transition metal oxides~\cite{MediciPRL2011,GeorgesARCMP2013}.

The most distinctive and intensively studied phenomenon induced by above two factors is the appearance of an orbital selective Mott (OSM) phase which is originally proposed to account for magnetic metallic state observed in Ca$_{2-x}$Sr$_x$RuO$_4$~\cite{AnisimovEPJB2002}. The OSM phase is characterized by coexistence of itinerant electrons and localized spins in same atomic subshell and is supposed to be the parent state for superconductivity in iron-based superconductors~\cite{MediciPRL2014}. The phase was recently detected in various iron chalcogenides~\cite{YiPRL2013,WangNaturecomm2014,PuPRB2016}. And within the OSM phase, it has been realized that magnetic and superconducting properties of iron-based superconductors can be successfully explained~\cite{YinPRL2011,YouPRL2011,Dainaturephys2012,TamPRL2015}. Moreover, it is known that the Kondo breakdown in heavy-fermion systems is conceptually identical to the quantum OSM phase~\cite{Neumannscience2007,VojtaJLTP2010,BeachPRB2011} and the momentum-space differentiation in doped cuprates can be viewed as the OSM phase in momentum-space~\cite{FerreroPRB2009}.

It is widely accepted that the Hund's coupling acts as a band decoupler through the suppression of orbital fluctuations~\cite{MediciPRB2011}, and the asymmetry among different orbitals due to various origins~\cite{MediciPRL2009,WernerPRL2007,KogaPRL2004,SongNJP2015,LeePRB2011} leads to Mott transitions separately taking place in each orbital at different interaction strength. As a consequence, localized and itinerant electrons coexist in a certain range of the on-site Coulomb repulsion.

Though above scenario derived from slave spin mean field~\cite{MediciPRB2011} approximation sounds plausible, a few fundamental questions remain unsolved. First, DMFT calculations with exact diagonalization (ED) as an impurity solver gave contradictory conclusions with respect to the ground state properties of the OSM phase~\cite{LiebschPRL2005,SunAPS2015}. While it was pointed out previously that the nature of the OSM phase is strongly dependent on whether the Hund's coupling is full or of Ising type~\cite{LiebschPRL2005} where the spin-flip and pair-hopping terms are absent, recent work indicates no qualitative difference between full and Ising Hund's couplings in the OSM phase~\cite{SunAPS2015}. Despite of the later work, the underlying physical quantity responsible for possible different nature of the OSM phase, i.e., the Fermi liquid behavior in the case of full Hund's coupling but non-Fermi liquid behavior in Ising limit~\cite{BiermannPRL2005}, is still unknown. Specifically, reminiscence of the study of kink energy scale in one-band Hubbard model~\cite{ByczukNaturePhy2007,VollhardtAP2012,RaasPRL2009,HeldPRL2013}, can the local spin susceptibility be a general quantity to distinguish the Fermi liquid from the non-Fermi liquid in the quantum OSM phase? Recently, a single kink energy scale along with a single maximum in the local spin susceptibility was found in the metallic state of a two-orbital Hubbard model with different bandwidth~\cite{GregerPRL2013}, indicating a close relation between energy scale of spin fluctuations present in the local spin susceptibility and the Fermi-liquid energy scale.

Second, a few recent studies based on slave spin mean field approximation supposed that the spin-flip and pair-hopping terms are negligible~\cite{YuRongPRL2013,FanfarilloPRB2015}, simply due to the fact that these terms are difficult to treat. Therefore, it is interesting to explore the validity of such an assumption. Third, DMFT calculations with different impurity solvers give opposite conclusions on whether there is an OSM phase in the two-orbital Hubbard model with Ising Hund's coupling at bandwidth ratio equal to $2$~\cite{KogaPB2005,KnechtPRB2005}. And finally, the order of the phase transitions is still under debate~\cite{KogaPRL2004,InabaPRB2006,MediciPRB2005,AritaPRB2005}.

In this paper, we will revisit the quantum OSM transitions based on a two-orbital Hubbard model with different bandwidth by employing combination of DMFT and ED~\cite{GeorgesRMP1996} at half filling. We will show that the order of OSM transitions can be clearly identified by the entanglement entropy and the inconsistency between DMFT calculations with different impurity solvers can be resolved. By analyzing local dynamical spin susceptibilities, we find that the presence of local spin fluctuations at low frequency is responsible for the Fermi liquid behavior in full Hund's coupling case while its absence indicates vanishing of the Kondo energy scale, leads to the absence of quasi-particle peak at the Fermi level and the appearance of non-Fermi liquid behavior, in Ising Hund's coupling case.

\section{ model and method }\label{maintext:modelmethod}

The two-orbital Hubbard model with narrow and wide bandwidth is defined as
\begin{eqnarray}\label{SCF:Hubbardmodel}
&H&=-\sum_{\langle ij\rangle \gamma \sigma} t^{\gamma}_{ij} c^{\dagger}_{i\gamma\sigma}c_{j\gamma\sigma} -\mu \sum_{i \gamma \sigma}n_{i\gamma\sigma}\label{eq:hamiltonian} \\
&+& U\sum_{i\gamma}n_{i\gamma\uparrow}n_{i\gamma\downarrow} +\Big(U'- J_z \Big)\sum_{i \sigma}n_{i a \sigma} n_{i b \sigma} \nonumber \\
&+& U'\sum_{i \sigma}n_{i a \sigma} n_{i b \bar{\sigma}} - J' \sum_{i} \left[S^{+}_{i a}S^{-}_{i b} +S^{-}_{i a}S^{+}_{i b}\right] \nonumber \\
&-& J^{p} \sum_{i}\left[c^{\dagger}_{i a \uparrow}c^{\dagger}_{i a  \downarrow}c_{i b \uparrow}c_{i b \downarrow}
+c^{\dagger}_{i b \uparrow}c^{\dagger}_{i b \downarrow}c_{i a \uparrow}c_{i a \downarrow},\right] \nonumber
\end{eqnarray}
where $t^{\gamma}_{ij}$ is hopping integral between nearest neighbor sites $i$ and $j$ denoted by $\langle ij\rangle$ in orbital $\gamma=N,W$. $U$, $U^{\prime }$ and $J_z$, $J'$, $J^{p}$ are the intra-orbital, inter-orbital Coulomb interaction and the Hund's coupling divided into Ising term, spin flip term, paring hopping term, respectively. $c^{\dagger}_{i\gamma\sigma}$ ($c_{i\gamma\sigma}$) creates (annihilates) an electron in orbital $\gamma$ of site $i$ with spin $\sigma$. $n_{i\gamma\sigma}=c^{\dagger}_{i\gamma\sigma}c_{i\gamma\sigma}$ is the occupation operator, while $S^{+}_{i\gamma}=c^{\dagger}_{i\gamma\uparrow}c_{i\gamma\downarrow}$ the spin operator. Here, $J'=J^{p}=J_z$ represents the case of full Hund's coupling while $J'=J^{p}=0$ stands for Ising Hund's coupling case. Both cases satisfy the condition of $U=U^{\prime }+2J_z$. The chemical potential of $\mu=U/2+U'-J_z/2$ is used to ensure half-filling condition in both bands.

We investigate the ground state properties of model~(\ref{eq:hamiltonian}) in the paramagnetic state by combination of DMFT and ED~\cite{GeorgesRMP1996,CaffarelPRL1994,ZhangPRB2007} where the two-orbital lattice model is mapped onto a two impurity Anderson model with each impurity coupled to $6$ discretized bath sites which are determined self-consistently through
\begin{eqnarray}
&&(g^{0}_{\gamma}\left( i\omega_n \right) )^{-1} - \Sigma_{\gamma} \left( i\omega_n \right) \\
&=&\left( \int \frac{d\epsilon \rho^{0}_{\gamma}\left( \epsilon \right) }{i\omega_n+\mu -\epsilon -\Sigma_{\gamma} \left( i\omega_n \right) } \right)^{-1}. \nonumber \label{selfconsistent}
\end{eqnarray}
Here, $i\omega_n$ is the Matsubara frequency, and $g^{0}_{\gamma}\left( i\omega_n \right) $ is the Weiss field where hybridization function of the impurity Anderson model is involved, while $\Sigma_{\gamma} \left( i\omega_n \right)$ is the local self-energy. We choose semielliptical density of states (DOS) as the noninteracting DOS for each orbital,
\begin{equation}
\rho^{0}_{\gamma}\left( \epsilon \right)=\frac{2}{\pi D_{\gamma}}\sqrt{1-(\frac{\epsilon}{D_{\gamma}})^2}
\end{equation}
where $D_{\gamma}$ is the half bandwidth. We use $D_{N}$ of narrower band as our unit of energy. In our calculations, we set an effective inverse temperature $\beta D_N=200$ which serves as a low-frequency cutoff. Unless specified otherwise, the case of bandwidth ratio of $D_{W}/D_{N}=2$ and $J_z=U/4$ is investigated, which is the most frequently studied case in the literature. (See Appendix \ref{sect:one} and \ref{sect:three} for details of the method and analyses of finite-size effect, respectively.)

\section{ results }\label{maintext:results}

\begin{figure}[htbp]
\includegraphics[width=0.48\textwidth]{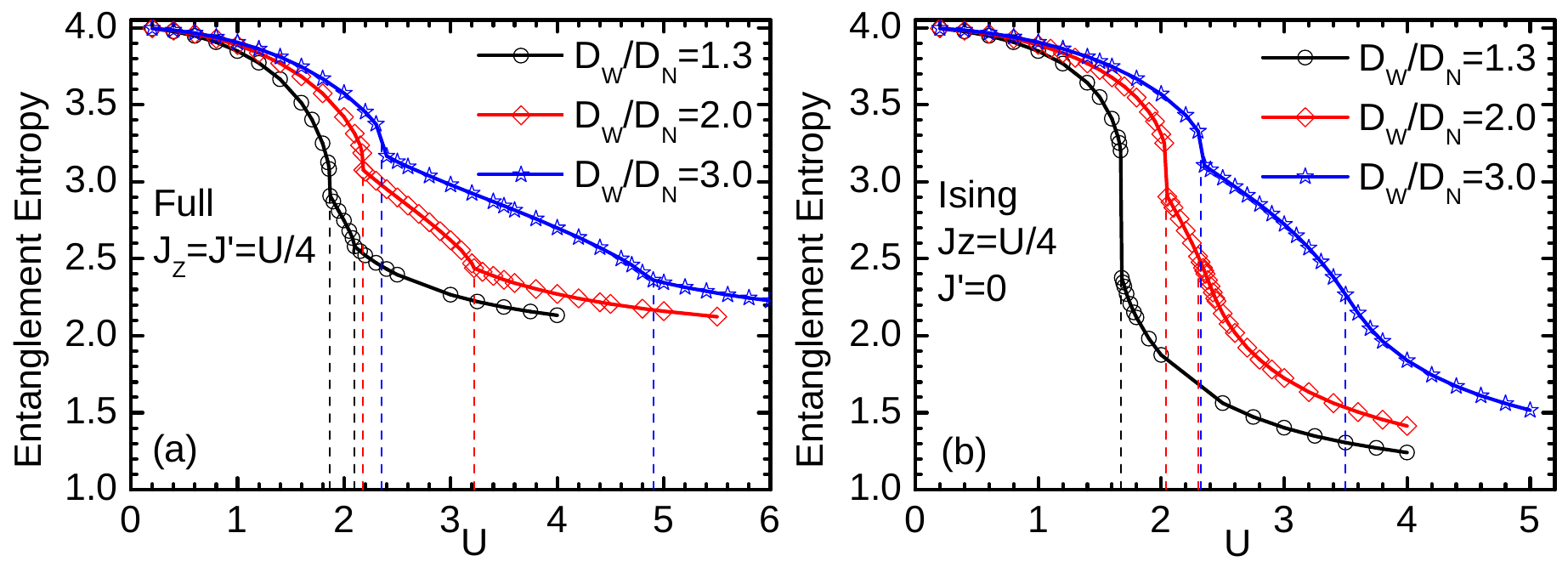}
\caption{(Color online) Entanglement entropy as a function of $U$ for three different bandwidth ratio in the case of full (a) and Ising (b) Hund's coupling. Dashed lines indicate the phase boundaries.}
\label{Fig:one}
\end{figure}

Fig.~\ref{Fig:one} shows for full and Ising Hund's coupling cases, the local entanglement entropy which measures the strength of quantum fluctuations and is sensitive to the presence of quantum phase transitions~\cite{AmicoRMP2008}, defined as $E_v=-\sum_{s=1}^{16} \lambda_s log_2 \lambda_s$. Here $\lambda_s$ is obtained from a reduced density matrix $\rho_i=Tr_i |\Psi_0\rangle\langle \Psi_0|=\sum_{s=1}^{16} \lambda_s |s\rangle\langle s|$, where $|\Psi_0\rangle$ is the ground state of two-orbital system and $Tr_i$ stands for tracing over all sites except the $i$th site. $|s\rangle$ denotes 16 possible local states at each site. (See Appendix \ref{subsect:energy} for advantages of using the entanglement entropy).

We find that for bandwidth ratio of $D_W/D_N=2$, there are two phase transitions in both full and Ising Hund's coupling cases. The phase boundaries are marked by dashed lines. While the first phase transition in each case is of first order, the second one is of different type. In full Hund's coupling case, a slope change can be observed at the second transition, indicating a discontinuity in the first derivative of entanglement entropy with respect to $U$ and occurence of a first-order phase transition, while a discontinuity can only be detected in second derivative of local entanglement entropy, indicating a second order phase transition in Ising Hund's coupling case. The results are fully consistent with those obtained from the derivatives of groundstate energy. (See Appendix \ref{subsect:energy})

The critical points are $U_{c1}\approx2.04$ and $U_{c2}\approx2.3$ in Ising Hund's coupling case, and $U_{c1}\approx2.18$ and $U_{c2}\approx3.22$ in full Hund's coupling case~\cite{comparison}. Similar calculations are done for bandwidth ratio of $D_W/D_N=3$ and $1.3$. Again, for $D_W/D_N=3$, intermediate phase is enlarged in full Hund's coupling case with $U_{c1}\approx2.35$ and $U_{c2}\approx4.9$, compared to that in Ising limit where $U_{c1}\approx2.33$ and $U_{c2}\approx3.5$. For $D_W/D_N=1.3$, intermediate phase vanishes and a direct metal to Mott insulator is detected in Ising limit with $U_{c}\approx1.68$, while intermediate phase is located between $U_{c1}\approx1.87$ and $U_{c2}\approx2.1$ in full Hund's coupling case. Obviously, there are remarkable differences between Ising and full Hund's coupling cases.

\begin{figure}[htbp]
\includegraphics[width=0.48\textwidth]{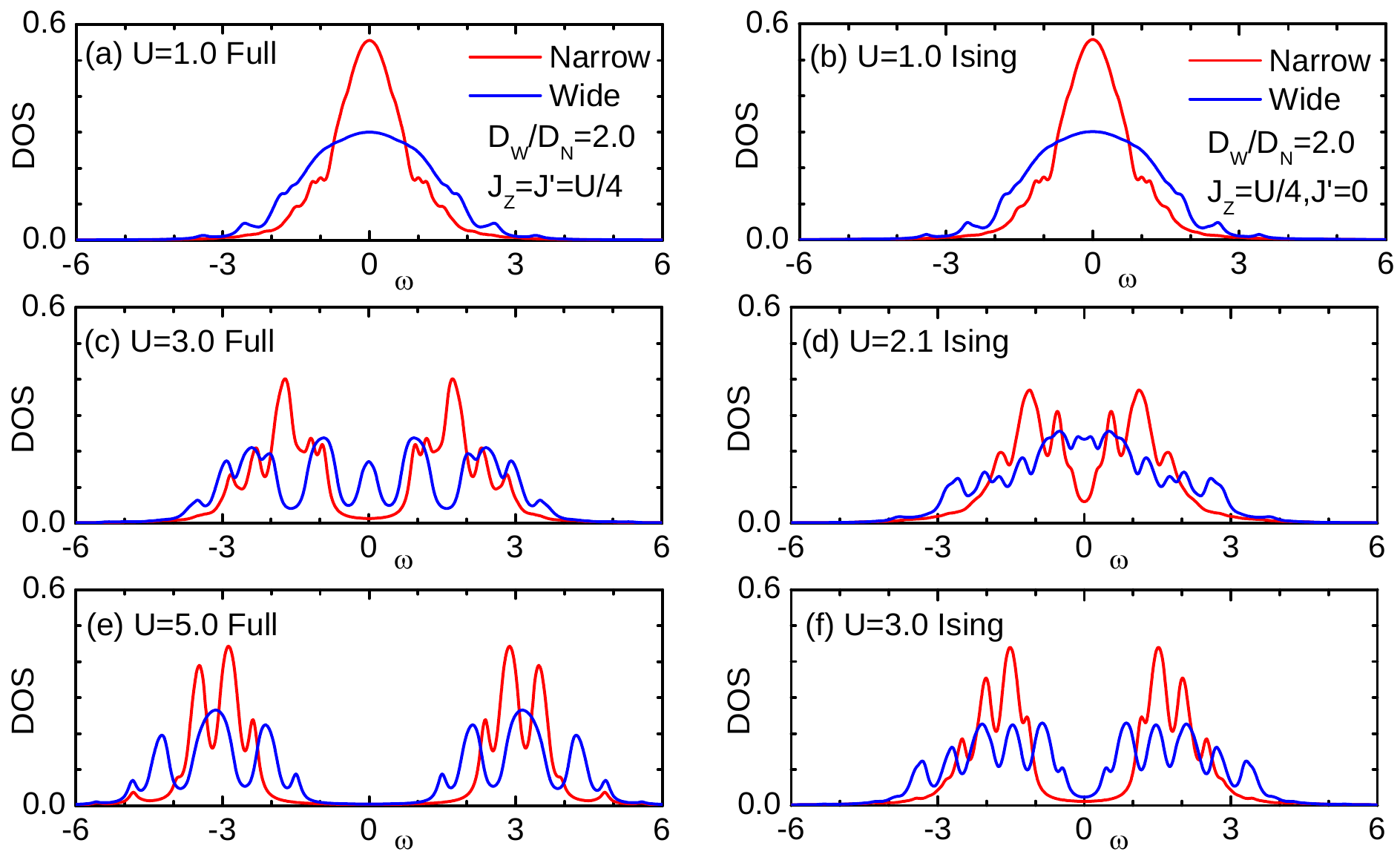}
\caption{(Color online) Orbital resolved densities of states in metallic states (a), (b), orbital selective Mott phases (c), (d) and Mott insulating states (e), (f) for full and Ising Hund's coupling cases. Here, a Lorentz broadening factor of $0.1$ is used.}
\label{Fig:two}
\end{figure}

The properties of the three phases in the two different cases are investigated in the following for $D_W/D_N=2$. From the plots of DOSs in Fig.~\ref{Fig:two}, we find that the systems undergo two consecutive phase transitions in both cases from metallic states with finite DOS at the Fermi level (Fig.~\ref{Fig:two} (a) and (b)) to Mott insulators with clear upper and lower Hubbard bands (Fig.~\ref{Fig:two} (e) and (f)) through intermediated OSM phases where one band is metallic while the other is insulating (Fig.~\ref{Fig:two} (c) and (d)) as a function of $U$. Furthermore, we can find in the OSM phases that, while a central peak preserves at the Fermi level in wide band of full Hund's coupling case (Fig.~\ref{Fig:two} (c)), a dip is present at $\omega=0$ in wide band of Ising Hund's coupling case (Fig.~\ref{Fig:two} (d)). The difference becomes even remarkable as the Lorentz broadening factor is extrapolated to zero. (See Appendix \ref{subsect:DOS} for more data and details).

\begin{figure}[htbp]
\includegraphics[width=0.48\textwidth]{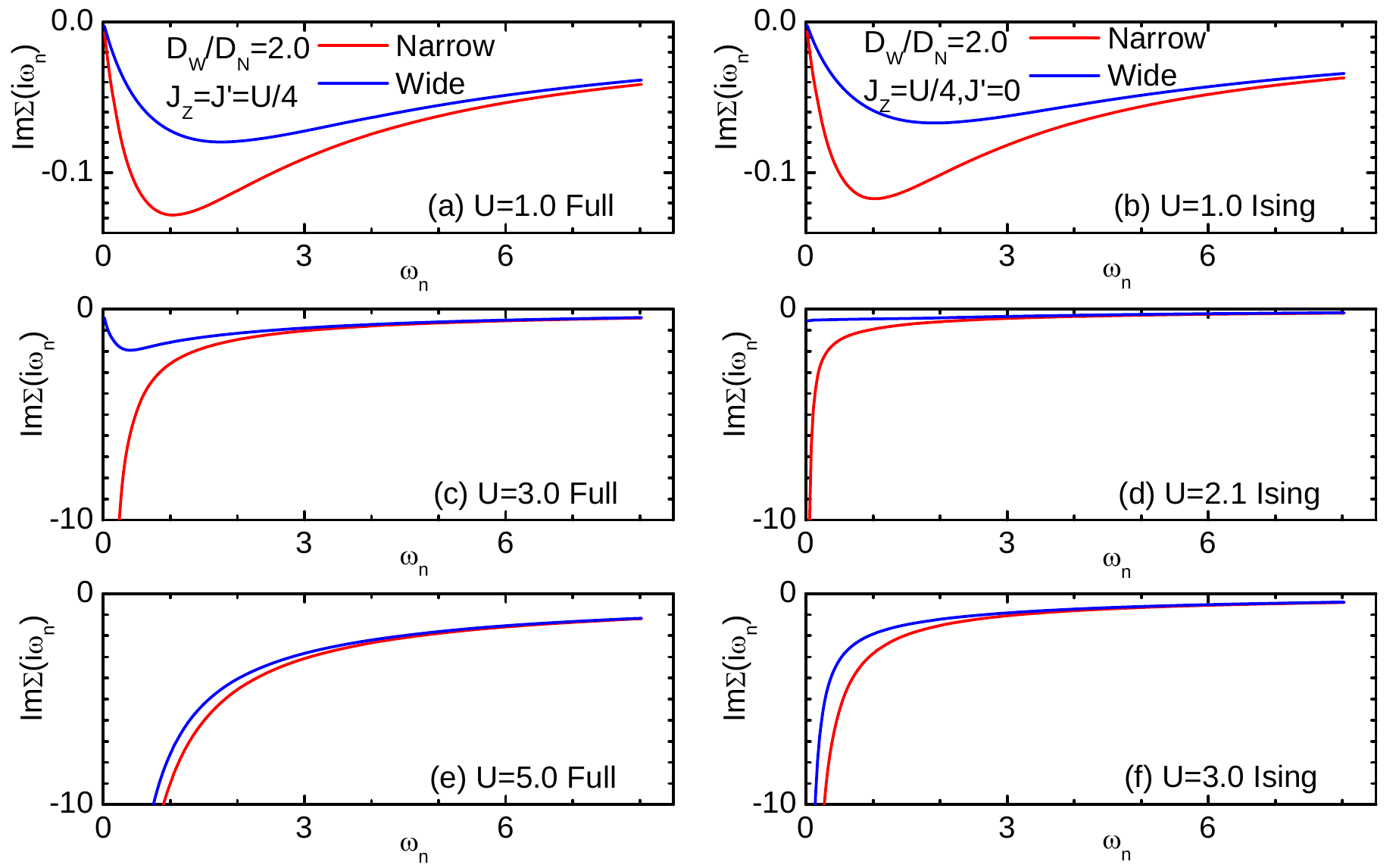}
\caption{(Color online) Orbital resolved imaginary parts of self-energies in metallic states (a), (b), orbital selective Mott phases (c), (d) and Mott insulating states (e), (f) for full and Ising Hund's coupling cases.}
\label{Fig:three}
\end{figure}

In order to determine the nature of the above three phases, we further investigate the imaginary part of self-energy in Matsubara frequency $Im\Sigma(i\omega_n)$ as depicted in Fig.~\ref{Fig:three}. In metallic phases (Fig.~\ref{Fig:two} (a) and (b)), $Im\Sigma(i\omega_n)$ of narrow and wide bands goes to zero as frequency goes to zero in each case, implying that both are the Fermi liquids. In insulating states (Fig.~\ref{Fig:two} (e) and (f)), all the quantities diverge proximity to zero frequency, which is a typical behavior of Mott insulator~\cite{ImadaRMP1998,GeorgesRMP1996}. However, in the intermediate OSM phases, $Im\Sigma(i\omega_n)$ of wide bands exhibits distinct behavior in different cases. It approaches zero as frequency goes to zero in full Hund's coupling case, $Im\Sigma(i\omega_n)$ remains finite in Ising Hund's coupling case, indicating presence of non-Fermi liquid where no long-lived quasiparticle exists at the Fermi level. (See Appendix \ref{subsect:selfenergyMatsu} for more data).

\begin{figure}[htbp]
\includegraphics[width=0.48\textwidth]{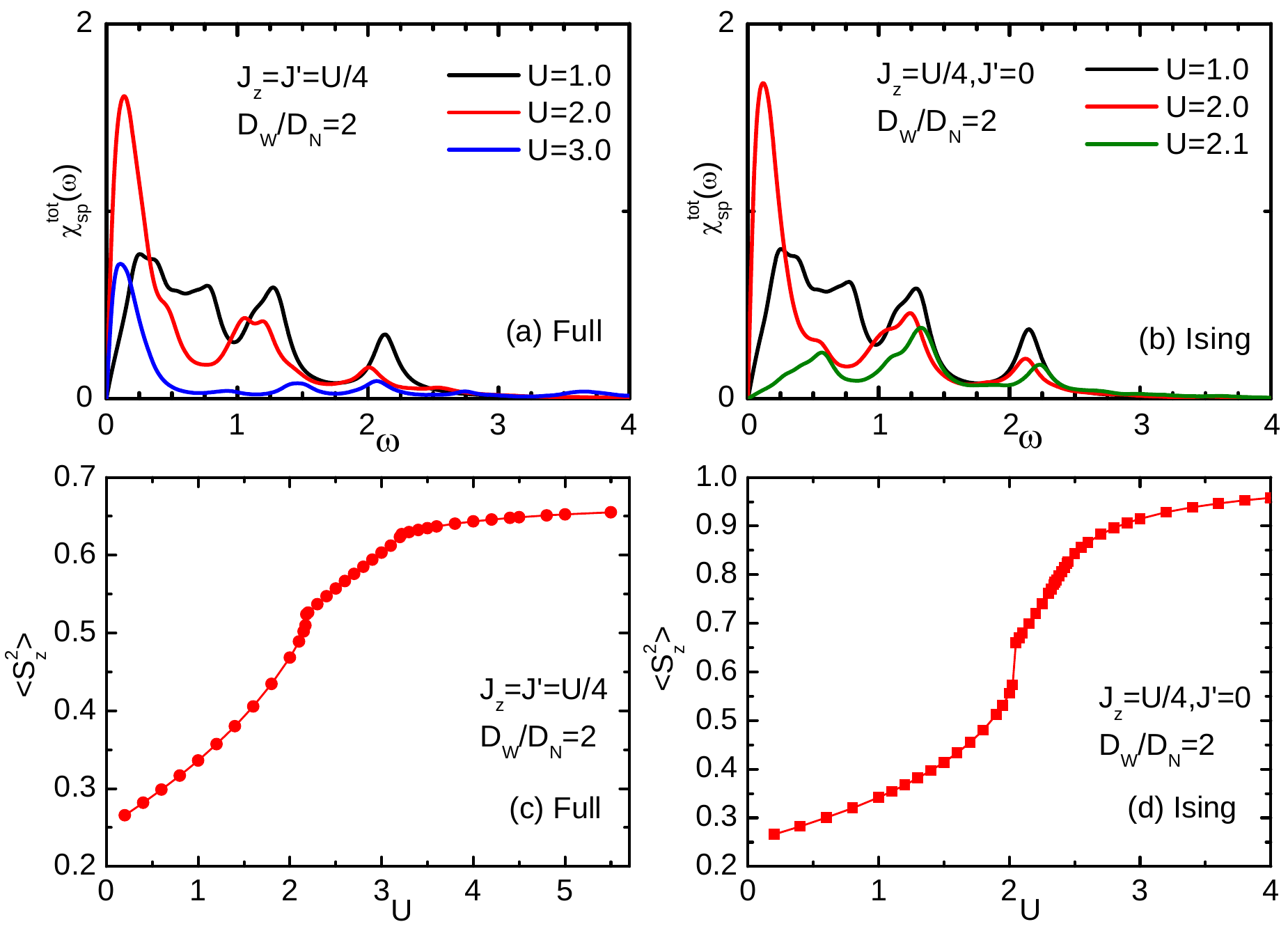}
\caption{(Color online) Total local dynamical spin susceptibilities in metallic and orbital selective Mott (OSM) phases for full (a) and Ising (b) Hund's coupling cases. Here, a Lorentz broadening factor of $0.1$ is used. (c) and (d) show $\langle S_z^2\rangle$ as a function of $U$ for full and Ising Hund's coupling cases, respectively.}
\label{Fig:four}
\end{figure}

It is interesting to explore the underlying physics of such a difference. Therefore, we calculate local dynamical spin correlations, defined as
\begin{equation}\label{maintext:chi}
\chi_{sp}^{\alpha \beta}(t)=-i\theta(t)\langle \Psi_0|[s_{\alpha}^{z}(t),s_{\beta}^{z}(0)]_{-}|\Psi_0\rangle,
\end{equation}
where $\theta(t)$ is a step function, $2s_{\alpha}^{z}(t)=n_{\alpha}^{\uparrow}-n_{\alpha}^{\downarrow}$ is $z$-component of local spin operator of $\alpha$ band at time $t$, $\Psi_0$ the groundstate, and $[,]_{-}$ denotes commutator of two operators. The total spin correlation is given by $\chi_{sp}^{tot}(t)=\sum_{\alpha,\beta} \chi_{sp}^{\alpha \beta}(t)$. After performing a Fourier transformation,
\begin{equation}
\chi^{tot}_{\xi}(z=\omega+i\eta)=-Im \int_{-\infty}^{\infty} e^{i z t} \chi^{tot}_{\xi}(t).
\end{equation}
we obtain local dynamical spin susceptibility as shown in Fig.~\ref{Fig:four}. Here, we only plot the imaginary parts since the real parts can be reproduced by Kramers-Kronig relations.

It is found that in each case, a strong peak at low frequency develops when $U$ approaches the critical value of first phase transition $U_{c1}$ as seen in Fig.~\ref{Fig:four} (a) and (b) from $U=1.0$ to $2.0$. The peak position gives a measure of the Kondo temperature and acts as the coherence scale below which the Fermi-liquid behavior sets in~\cite{VollhardtAP2012}. As $U$ further increases, the OSM phase appears. The spin excitations becomes completely different between full and Ising Hund's coupling cases (see $U=3.0$ and $2.1$ in Fig.~\ref{Fig:four} (a) and (b), respectively). While the peak at low frequency entirely vanishes in Ising Hund's coupling case, it moves to even lower frequency with reduced intensity in full Hund's coupling case. Such a difference can be understood from strong interaction limit. As shown in Fig.~\ref{Fig:four} (c) and (d), $\langle S_z^2 \rangle$ asymptotically goes to $2/3$ and $1$ at large $U$ limit in full and Ising Hund's coupling cases, indicating spin triplet and doublet states, respectively. Since there are three degenerate states ($S_z=\pm1,0$) in local Hilbert space in full Hund's coupling case, flipping a spin may cost low energy of the order of $D_W^2/(U'-J_z-\mu)$ through a virtual process like $|\uparrow,\uparrow\rangle \downarrow$$\rightarrow$$|\uparrow,0\rangle \uparrow\downarrow$$\rightarrow$$1/\sqrt{2}(|\uparrow,\downarrow\rangle+|\downarrow,\uparrow\rangle)\uparrow$, resulting in a peak of spin excitations at low frequency. Here, arrows inside $|,\rangle$ represent for spin states of electrons on two orbitals of one site, and those outside $|,\rangle$ denote spin states of electrons on rest sites that are involved in the process. The reduction of the intensity of low-energy spin excitations across the transition from metal to the OSM phase is due to the fact that in metallic state, both narrow and wide bands provide channels for hybridizations between neighboring sites, while in the OSM phase, only wide band allows electrons to hop. In contrast, there are only two degenerate states ($S_z=\pm1$) in Ising Hund's coupling case. Therefore, no low-energy channel through virtual process exists and flipping a spin from $S_z=\pm1$ ($|\uparrow,\uparrow\rangle$ or $|\downarrow,\downarrow\rangle$) to $S_z=0$ ($|\uparrow,\downarrow\rangle$) state simply costs energy of the order of $J_z$. Such spin excitations are corresponding to the first peak close to $\omega=0$ as seen in Fig.~\ref{Fig:four} (b) for $U=2.1$.
(See Appendix \ref{subsect:dynamical} for more data).

\section{ discussion }

From above comparison of local dynamical spin susceptibilities in different OSM phases, it is obvious that disappearance of low-energy local spin fluctuations leads to vanishing of Kondo or coherence energy scale and emergence of non-Fermi liquid. Further analyzing orbital resolved spin susceptibilities, we find that in the Fermi-liquid OSM phase the spin excitations are both gapless in wide and narrow bands even though narrow band is insulating, while in the non-Fermi-liquid OSM phase, spin excitations are gapped in narrow band but remain gapless in wide band. (See Fig.~\ref{Fig:apdx:6bathspinSUS} in Appendix) These indicate that although charge degrees of freedom between orbitals are decoupled in both OSM phases~\cite{MediciPRB2011}, the spin degrees of freedom behave distinctly, dependent on whether the coupling between orbitals is full or of Ising type. If full, the spins of different orbitals are strongly coupled with each other. Otherwise, the spins are decoupled between orbitals and the spin in narrow band acts as an effective magnetic field relative to that in wide band. Therefore, the widely accepted scenario for the OSM phase~\cite{MediciPRB2011} should be revised.

Finally, we should point out that although we only study full and Ising Hund's coupling cases, the results of the cases with $J'<J_z$ should be similiar to those in Ising limit. This can be understood from renormalization group flow of the ferromagnetic Kondo impurity model~\cite{AndersonJPC1970}. For a metallic band, it predict that spin flip term $J'$ is renormalized to zero but the Ising term $J_z$ to a finite value. For an insulating band, since the system is gapped, $J'<J_z$ becomes irrelevant for low-energy properties and the Ising term is renormalized to a finite value. Above analyses are further corroborated by numerical calculations on a ferromagnetic Kondo lattice model including onsite Coulomb repulsion $U$~\cite{BiermannPRL2005,CostiPRL2007}. Therefore, our results indicate that quantum non-Fermi liquid, which is characterized by vanishing of low-energy local spin fluctuations, may generally exist in many correlated multi-orbital materials if those are in OSM phases, since Hund's coupling is usually anisotropic due to noncubic octahedral distortions.

\section{Conclusion}

In conclusion, we find that local entanglement entropy can be used to determine the critical values and the nature of OSM phase transitions efficiently and thus can resolve existing contradictions. We reveal that the absence of low-energy local spin fluctuations in the OSM phase is responsible for the appearance of non-Fermi liquid at zero temperature. Our study indicates that the OSM transition happens when orbitals are decoupled in charge channel, while the nature of the OSM phase is determined by the fact of whether the orbitals are decoupled in spin channel, which is different from the widely accepted senario for the OSM phase~\cite{MediciPRB2011}.

In fact, absence of low-energy local spin fluctuations may also be a general origin for the non-Fermi liquid observed in one-band Hubbard model. As short-range antiferromagnetic order is taken into account on non-frustrated lattice, for example the study of Mott transition by cluster extension of DMFT~\cite{ZhangPRB2007,ParkPRL2008}, the low-energy local spin fluctuations have to be suppressed due to the development of inter-site spin-spin correlations as a function of $U$, and therefore non-Fermi liquid appears in the vicinity of Mott transition. Our findings on the origin of quantum non-Fermi liquid may also be applicable to account for non-Fermi-liquid state in heavy Fermion metals with crystal field anisotropy as it was accepted that the Kondo-breakdown and the quantum OSM phase transitions are conceptually identical~\cite{VojtaJLTP2010}.

Moreover, it is interesting to go beyond the local approximation to the electronic correlations, a limitation of single-site DMFT, and consider the effect of nonlocal correlations as well as various ordered states on the OSM phase, since various antiferromagnetic fluctuations were reported in iron-based superconductors. Recently, pioneer works in one dimensional systems have revealed a lot of novel results related to the OSM phase after spatial fluctuations are fully taken into account~\cite{RinconPRL2014,RinconPRB2014,LiPRB2016}.

\section*{Acknowledgement}\label{Acknowledgement}

This work is supported by National Natural Science Foundation of China (Nos. 11474217, 11774258, 11174219). H.Q. Lin acknowledges support from NSAF U1530401 and computational resource from the Beijing Computational Science Research Center.

\appendix

\section{introduction to dynamical mean field theory with exact diagonalization\label{sect:one}}
In the framework of dynamical mean field theory (DMFT), the lattice model is mapped onto an Anderson impurity model. The mapping is exact in infinite dimension~\cite{GeorgesRMP1996}. According to this mapping, the local Green's function and the self-energy of lattice model must be identical to those of the Anderson impurity model, which serves as a self-consistent condition.  When exact diagonalization (ED) is employed as an impurity solver~\cite{GeorgesRMP1996,CaffarelPRL1994} as is used in this paper, the crucial step is to establish an effective Anderson impurity model with discretized bath and optimized bath parameters, with which the Weiss field of the original lattice model can be reproduced by the noninteracting Green's function of the discretized Anderson impurity model as precisely as possible, in each self-consistent loop.
\par
In the following, we will describe above process based on the two-orbital Hubbard model (\ref{SCF:Hubbardmodel}) which is defined in the Sec.~\ref{maintext:modelmethod}. The local Green's function of the two-orbital Hubbard model (\ref{SCF:Hubbardmodel}) is calculated as
\begin{align}\label{SCF:local_Green}
G_{\gamma}(i\omega_n)=\int^{+\infty}_{-\infty}
\frac{\rho_{\gamma}(\epsilon)d\epsilon}{i\omega_n+\mu-\Sigma_{\gamma}(i\omega_n)-\epsilon},
\end{align}
where $\gamma=\text{N, W}$ is band index with N and W denoting narrow and wide band, respectively. $\rho_{\gamma}(\epsilon)=2\sqrt{1-(\frac{\epsilon}{D_{\gamma}})^2}/\pi{D}_{\gamma}$ is the noninteracting density of states of $\gamma$ band of the two-orbital Hubbard model (\ref{SCF:Hubbardmodel}) on Bethe lattice, where $D_{\gamma}$ is the half bandwidth. $\mu$ is chemical potential and $\Sigma_{\gamma}(i\omega_n)$ is the local self-energy of $\gamma$ band. Here, $\omega_n$ is the Matsubara frequency. The Weiss field $g_{0\gamma}(i\omega_n)$ can be obtained through the Dyson equation as
\begin{align}\label{SCF:weiss_field}
g^{-1}_{0\gamma}(i\omega_n)=G_{\gamma}^{-1}(i\omega_n)+\Sigma_{\gamma}(i\omega_n).
\end{align}
\par
On the other hand, the noninteracting Green's function $\widehat{g}_{0\gamma}(i\omega_n)$ of the corresponding two-orbital Anderson impurity model can be written as
\begin{align}\label{SCF:weiss_impurity}
\widehat{g}^{-1}_{0\gamma}(i\omega_n)=
i\omega_n+\mu-\sum_{k}\frac{V^{*}_{k\gamma}V_{k\gamma}}{i\omega_n-\epsilon_k}
\end{align}
where $\epsilon_k$ is the bath energy, $V_{k\gamma}$ denotes the hybridization between bath and the local orbitals on the impurity. The corresponding two-orbital Anderson impurity model reads
\begin{align}\label{SCF:impurity_model}
H_{\text{eff}}&=\sum_{k\sigma}\epsilon_{k}C^{\dag}_{k\sigma}C_{k\sigma}
+\sum_{k\gamma\sigma}\left[V_{k\gamma}C^{\dag}_{k\sigma}d_{\gamma\sigma}+
V^*_{k\gamma}d^{\dag}_{\gamma\sigma}C_{k\sigma}\right] \nonumber\\
&+U\sum_{\gamma}n_{\gamma\uparrow}n_{\gamma\downarrow}+
\left(U^{\prime}-J_z\right)\sum_{\sigma}n_{N\sigma}n_{W\sigma} \nonumber\\
&+U^{\prime}\sum_{\sigma}n_{N\sigma}n_{W\bar\sigma}
-J^{\prime}\left[S^{+}_NS^{-}_W+S^{-}_NS^{+}_W\right] \nonumber\\
&-J^p\left[C^{\dag}_{N\uparrow}C^{\dag}_{N\downarrow}C_{W\uparrow}C_{W\downarrow}+
           C^{\dag}_{W\uparrow}C^{\dag}_{W\downarrow}C_{N\uparrow}C_{N\downarrow}\right],
\end{align}
with $U$, $U^{\prime}$ and $J_z$, $J^{\prime}$, $J^{p}$ the same meaning as their counterparts of the original two-orbital Hubbard model (\ref{SCF:Hubbardmodel}) on Bethe lattice.
\par
In order to feasibly solve the effective two-orbital Anderson impurity model (\ref{SCF:impurity_model}) with ED, we have to use finite number of bath sites to optimally fit the continuous bath. In this paper, we use 6 discrete bath sites coupled to each orbital, i.e., 12 bath sites in total.
\par
In practice, we start from an initial set of bath parameters $\{\epsilon_k,V_{k\gamma}\}$ and calculate the impurity Green's function $\widehat{G}_{\gamma}(i\omega_n)$ using ED and the noninteracting Green's function $\widehat{g}_{0\gamma}(i\omega_n)$, based on the effective two-orbital Anderson impurity model (\ref{SCF:impurity_model}). Then the local self-energy $\Sigma_{\gamma}(i\omega_n)$ can be obtained through the Dyson equation as
\begin{align}\label{SCF:self_enegy}
\Sigma_{\gamma}(i\omega_n)=\widehat{g}^{-1}_{0\gamma}(i\omega_n)-\widehat{G}^{-1}_{\gamma}(i\omega_n)
\end{align}
Having obtained the local self-energy $\Sigma_{\gamma}(i\omega_n)$, one can use
Eq.~(\ref{SCF:local_Green}) to obtain the local Green's function of the original two-orbital Hubbard model (\ref{SCF:Hubbardmodel}) and a new estimate for the Weiss field $g_{0\gamma}(i\omega_n)$ is given by Eq. (\ref{SCF:weiss_field}). When the new Weiss field $g_{0\gamma}(i\omega_n)$ is obtained, we must search a new set of bath parameters $\{\epsilon_k,V_{k\gamma}\}$, such that the discrete version of the noninteracting impurity Green's function $\hat{g}_{0\gamma}(i\omega_n)$ gives a best fit to the new Weiss field $g_{0\gamma}(i\omega_n)$. In order to fulfill this requirement, we employ the conjugate gradient method to minimize a cost function which is defined as
\begin{align}
\chi^2=\frac{1}{N_{\max}}\sum^{N_{\max}}_{n=1}W(\omega_n)
\sum_{\gamma}|g^{-1}_{0\gamma}(i\omega_n)-\hat{g}^{-1}_{0\gamma}(i\omega_n)|^2
\end{align}
where the Matsubara frequency is written as $\omega_n=(2n+1)\pi/\beta$ with a fictitious temperature $\beta D_N=200$, which serves as a low frequency cutoff. $N_{\max}=256$ is the upper limit of the summation, and $W(\omega_n)=1/\omega^2_n$ is a frequency dependent weight used to obtain a best fit for low frequency part of $g_{0\gamma}(i\omega_n)$. After the new effective two-orbital Anderson impurity model (\ref{SCF:impurity_model}) is built in term of the new set of bath parameters $\{\epsilon_k,V_{k\gamma}\}$ produced by the conjugate gradient method, the impurity solver ED is again used to solve the new discretized two-orbital Anderson impurity model (\ref{SCF:impurity_model}) and a new local self-energy $\Sigma_{\gamma}(i\omega_n)$ is calculated. Such iterations continue until convergence is reached, where the difference $\Delta_{g_0}$ between new Weiss field $g^{\text{new}}_{0\gamma}(i\omega_n)$ and the old Weiss field $g^{\text{old}}_{0\gamma}(i\omega_n)$ is less than a tolerance of $10^{-6}$. The difference $\Delta_{g_0}$ is defined as
\begin{align}
\Delta_{g_0}=\max\{|g^{\text{new}}_{0\gamma}(i\omega_n)-g^{\text{old}}_{0\gamma}(i\omega_n)|^2_{\gamma=N,W}\}
\end{align}


\section{Details of our results\label{sect:two}}
In this work, the two-orbital Hubbard model on the Bethe lattice with infinite connectivity $z$ as defined in the Sec.\ref{maintext:modelmethod} is investigated by dynamical mean field theory in combination with exact diagonalization~\cite{GeorgesRMP1996,CaffarelPRL1994}. Both full and Ising Hund's coupling are studied at different bandwidth ratios, namely $D_{W}:D_{N}=1.3:1$, $D_{W}:D_{N}=2:1$ and $D_{W}:D_{N}=3:1$. It shows that there are distinct differences between full and Ising coupling. Firstly, although the OSM phase is observed in both full and Ising Hund's coupling cases at bandwidth ratios of $D_{W}:D_{N}=2:1$ and $D_{W}:D_{N}=3:1$, the region of the OSM phase for full Hund's coupling is much wider than that for Ising Hund's coupling. Besides, at the bandwidth ratio of $D_{W}:D_{N}=1.3:1$, the OSM phase takes place in the case of full Hund's coupling, while only a single Mott transition is observed in the case of Ising Hund's coupling. Secondly, the asymptotic behaviors of DOS and imaginary part of self-energy in Matsubara frequency proximity to the Fermi level in the OSM phase indicate that a Fermi liquid state appears in the wide band for full Hund's coupling while a non-Fermi liquid state is present for Ising Hund's coupling. Furthermore, the low-energy local spin excitations persist in the OSM phase for full Hund's coupling, while those disappear for Ising Hund's coupling. Through analysing the results presented above, we conclude that the distinct behavior of the low-energy local spin excitations is responsible for the remarkable difference in the OSM phase between full and Ising Hund's coupling. In the following, we supplement a detailed comparison between the cases of full and Ising Hund's coupling, especially in the OSM phase, at a typical bandwidth ratio of $D_{W}:D_{N}=2:1$.

\subsection{DOS in the OSM phase\label{subsect:DOS}}
\begin{figure}[htbp]
\includegraphics[width=0.48\textwidth]{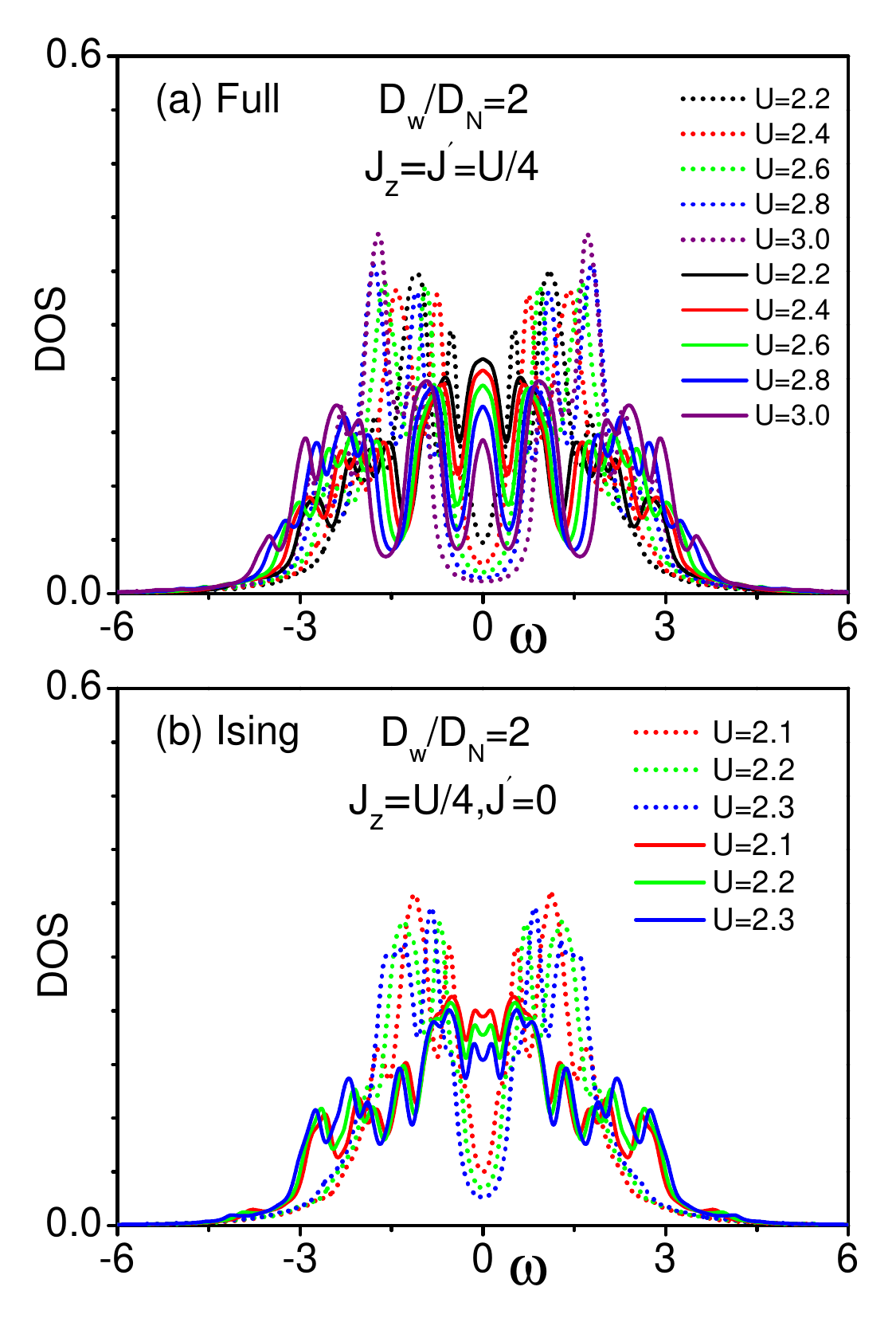}
\caption{(Color online) The U dependence of DOS in the OSM phase for the cases of full (a) and Ising (b) Hund's coupling. Solid lines denote wide band and dashed lines represent narrow band. Here, a Lorentz broadening factor of $0.1$ is used.}
\label{Fig:apdx:6bathDOS}
\end{figure}
In Fig.~\ref{Fig:apdx:6bathDOS}, we plot the $U$ dependence of DOS in the OSM phase for both full and Ising Hund's coupling. It can be seen that the narrow band is insulating while the wide band remain metallic, indicating the presence of the OSM phase, for all values of $U$ in both cases. Moreover, a central peak at Fermi level is present in wide band for full Hund's coupling,  indicating the presence of the Fermi liquid, while a dip appears in wide band for Ising Hund's coupling, implying the occurrence of non-Fermi liquid.

\begin{figure}[htbp]
\includegraphics[width=0.48\textwidth]{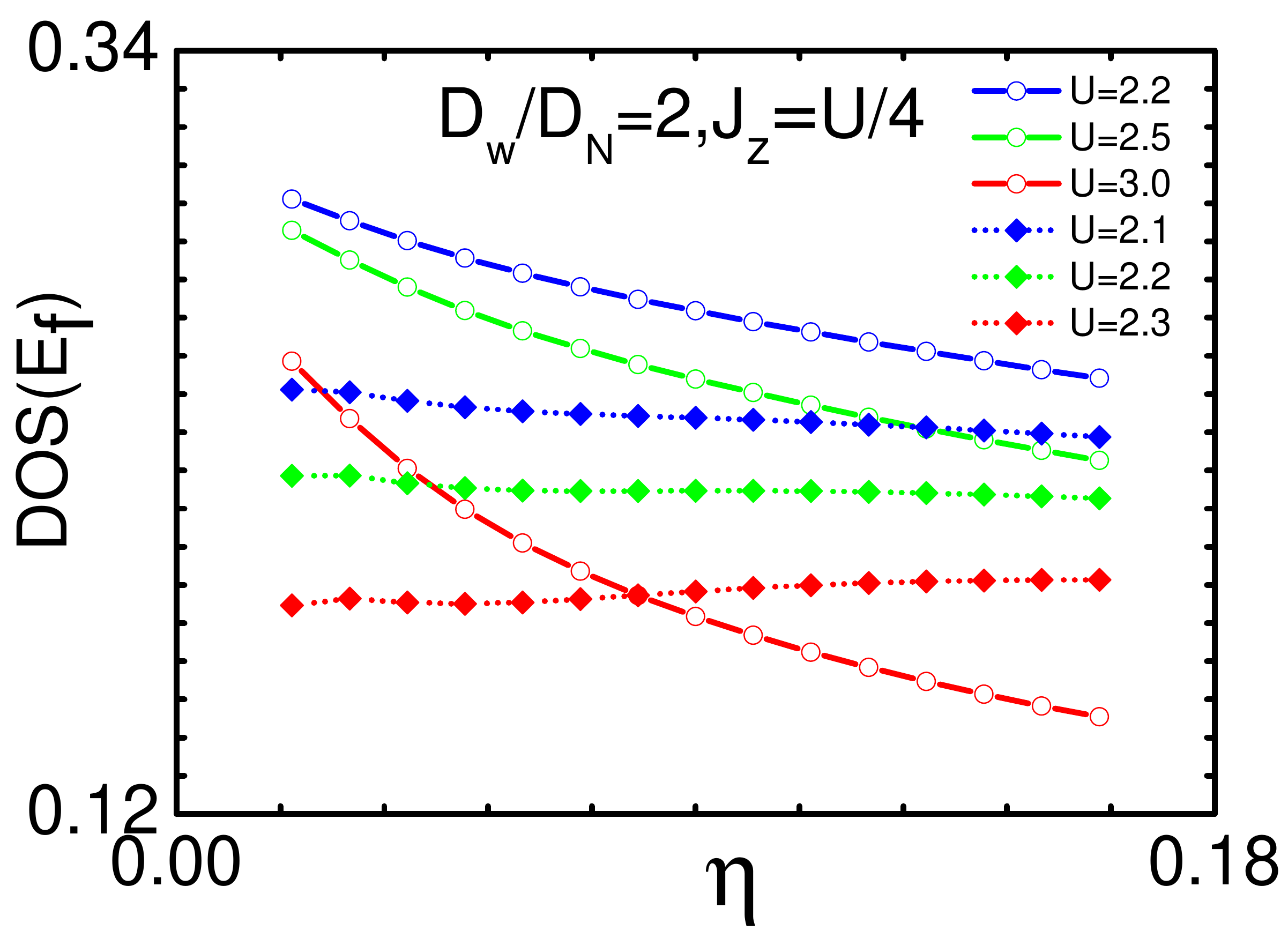}
\caption{(Color online) The evolution of DOS of wide band at the Fermi level as a function of the lorentz broadening factor $\eta$ in full and Ising Hund's coupling cases. solid lines with empty circle denote full Hund's coupling, dashed lines with filled diamond denote Ising Hund's coupling. }
\label{Fig:apdx:DOSlorentz}
\end{figure}

The difference becomes even remarkable when the lorentz broadening factor $\eta$ is extrapolated to zero as depicted in Fig.~\ref{Fig:apdx:DOSlorentz}. For full Hund' coupling, the DOS at the Fermi level increases as the broadening factor $\eta$ approaches zero, but that almost keeps constant for Ising Hund's coupling. The results suggest that the scattering rate or inverse of the life time of quasiparticles is dominated by the imaginary part of self-energy for Ising Hund's coupling, while it is controlled by the lorentz broadening factor $\eta$ for full Hund's coupling. This is again a clear evidence that for Ising Hund's coupling, there is no long-lived quasiparticle, irrespective of how the broadening factor is chosen, while for full Hund's coupling, quasiparticles have infinite life time as the lorentz broadening factor $\eta$ goes to zero. These conclusions can be confirmed by the imaginary part of self-energy in Matsubara frequency as presented in  Appendix~\ref{subsect:selfenergyMatsu}.
\subsection{imaginary part of the self-energy in Matsubara frequency\label{subsect:selfenergyMatsu}}
\begin{figure}[htbp]
\includegraphics[width=0.48\textwidth]{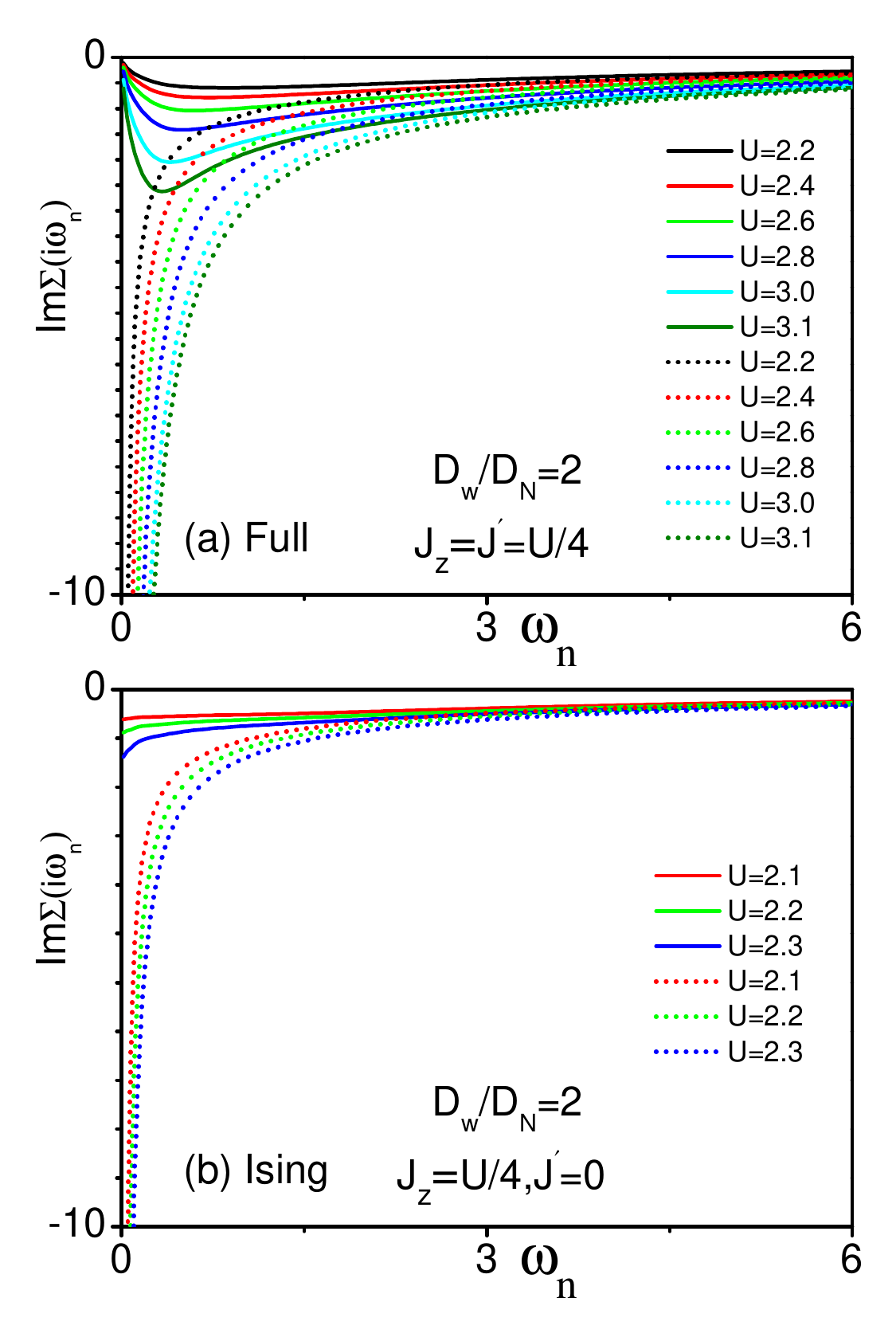}
\caption{(Color online) The $U$ dependence of imaginary part of self-energy in the Matsubara frequency $Im\Sigma(i\omega_n)$ in the OSM phase in the case of full (a) and Ising (b) Hund's coupling. Solid lines denote wide band and dashed lines represent narrow band. }
\label{Fig:apdx:6bathSelfEg}
\end{figure}
The $U$ dependence of imaginary part of self-energy on the Matsubara axis $Im\Sigma(i\omega_n)$ in the OSM phase is illustrated in Fig.~\ref{Fig:apdx:6bathSelfEg}. For both full (Fig.~\ref{Fig:apdx:6bathSelfEg} (a)) and Ising (Fig.~\ref{Fig:apdx:6bathSelfEg} (b)) Hund's coupling, $Im\Sigma(i\omega_n)$ of narrow band diverges as the Matsubara frequency $\omega_n$ is extrapolated to zero for all values of $U$, indicating that the narrow band becomes a Mott insulator. However, $Im\Sigma(i\omega_n)$ of wide band exhibits different behavior in different cases. It goes to zero as Matsubara frequency $\omega_n$ goes to zero in the case of full Hund's coupling, implying the presence of Fermi liquid, while it approaches a finite value in the low-frequency limit in the case of Ising Hund's coupling, suggesting finite scattering rate and no long-lived quasiparticles at the Fermi level, and the appearance of non-Fermi liquid. The results of imaginary part of self-energy in the Matsubara frequency are consistent with those of DOS in the OSM phase as presented in Appendix~\ref{subsect:DOS}.
\subsection{clarification of effective mass for Ising Hund's coupling\label{subsect:effectivemass}}
\begin{figure}[htbp]
\includegraphics[width=0.48\textwidth]{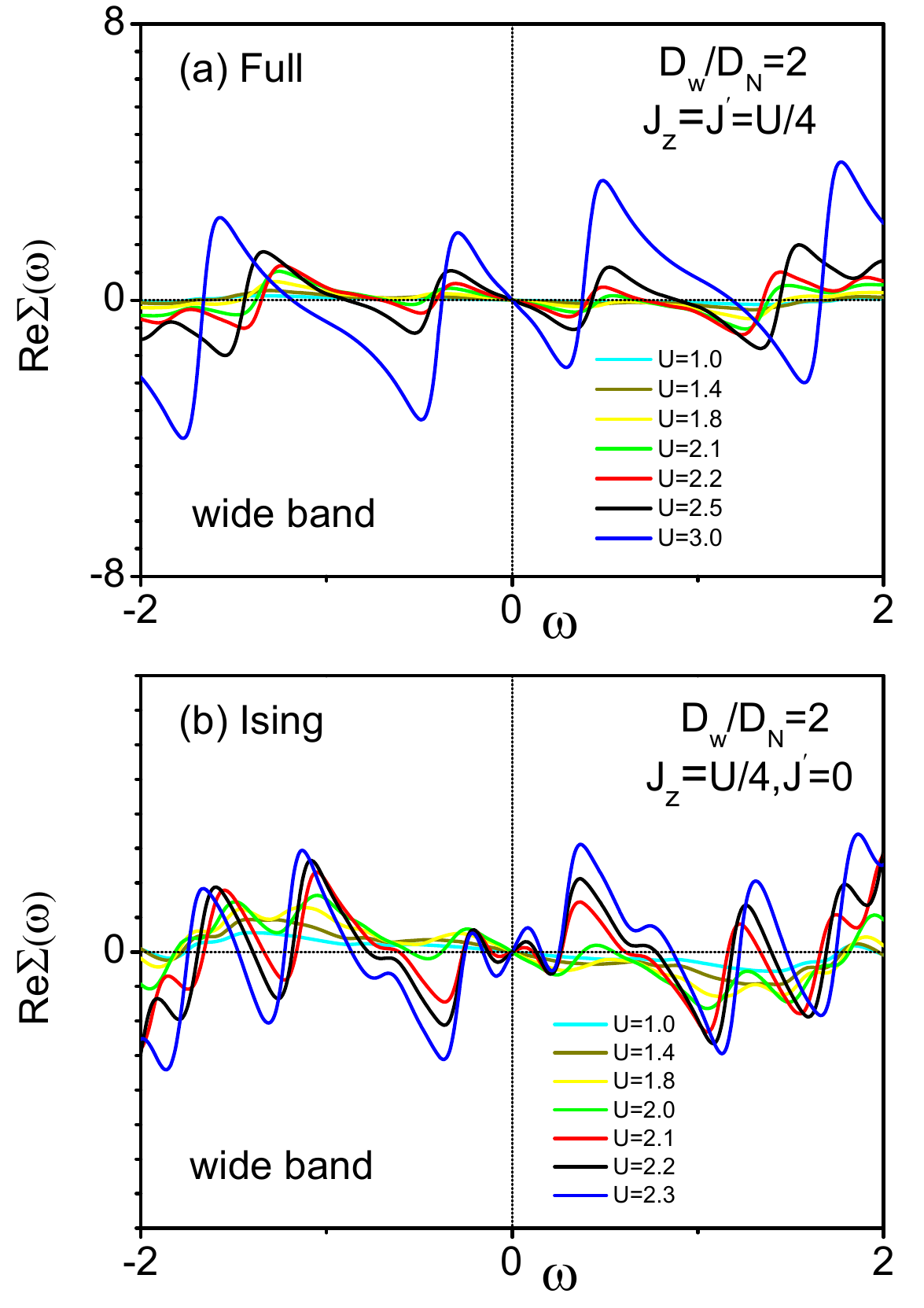}
\caption{(Color online) The $U$ dependence of real part of self-energy in real frequency $Re\Sigma(\omega)$ of wide band from metallic to the OSM phase for full (a) and Ising (b) Hund's coupling. Here, a Lorentz broadening factor of $0.1$ is used.}
\label{Fig:apdx:RealSelfEg}
\end{figure}
Evolution of the real part of self-energy of wide band in real frequency $Re\Sigma(\omega)$ from metallic to the OSM phase as a function of $U$ is displayed in Fig.~\ref{Fig:apdx:RealSelfEg}. From the behavior of $Re\Sigma(\omega)$ close to $\omega=0$, it is found that there is a remarkable difference in the OSM phase between full (Fig.~\ref{Fig:apdx:RealSelfEg} (a)) and Ising (Fig.~\ref{Fig:apdx:RealSelfEg} (b)) Hund's coupling. The slope of real part of self-energy keeps negative in the vicinity of the Fermi level in both metallic and the OSM phases for full Hund's coupling, while it changes abruptly from negative in metallic phase to positive in the OSM phase near the Fermi level for Ising Hund's coupling. The positive slope is a characteristic feature for non-Fermi liquid. By the definition of the effective mass where
\begin{align}\label{SCF:effectivemass}
m^{*}/m_{0}=(1-\frac{\partial Re\Sigma_{\gamma}(\omega)}{\partial\omega}\mid_{\omega=0}),
\end{align}
the effective mass of a correlated electron in non-Fermi-liquid phase will be oddly smaller than the mass of a bare electron. That's the reason why we do not use the effective mass to investigate the OSM transition, since there is no long-lived quasiparticle in non-Fermi-liquid phase and the effective mass is not well defined. Most of the previous studies based on quantum Monte Carlo solver~\cite{KnechtPRB2005,JakobiPRB2009} and slave spin method~\cite{YuRongPRL2013,FanfarilloPRB2015} were not aware of this crucial point, simply due to the fact that slave spin method does not take into account the dynamical fluctuations  and quantum Monte Carlo solver treats the effective mass approximately as
\begin{align}\label{SCF:effectivemass:apdx}
m^{*}/m_{0} \approx 1-\frac{Im\Sigma_{\gamma}(i\omega_0)}{\omega_0},
\end{align}
\subsection{dynamical spin susceptibility\label{subsect:dynamical}}
\begin{figure}[htbp]
\includegraphics[width=0.48\textwidth]{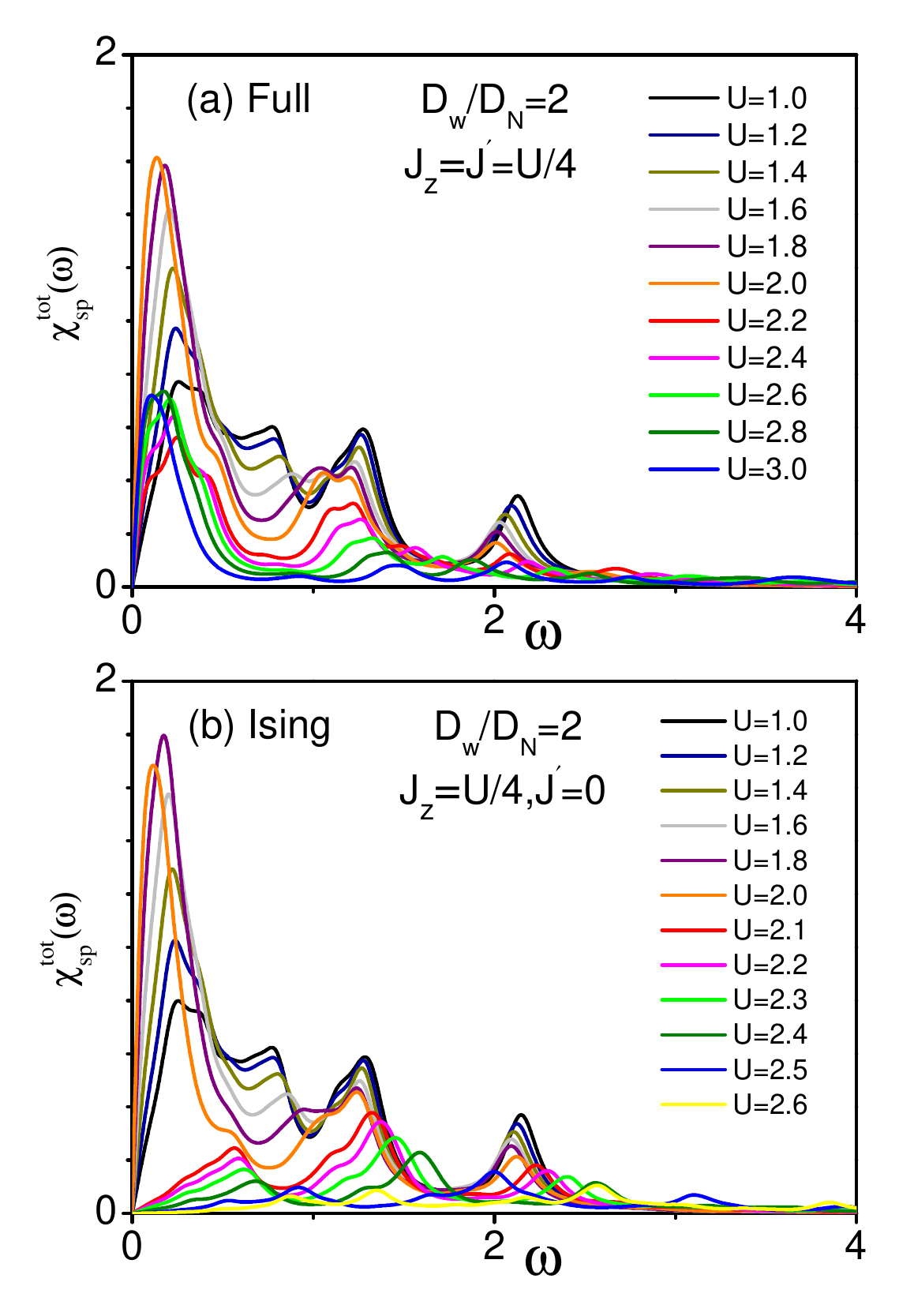}
\caption{(Color online) The evolution of dynamical spin susceptibility $\chi^{\text{tot}}_{\text{sp}}(\omega)$ as a function of $U$ in the case of full (a) and Ising (b) Hund's coupling. Here, a Lorentz broadening factor of $0.1$ is used.}
\label{Fig:apdx:6bathChi}
\end{figure}
We display the evolution of dynamical spin susceptibility $\chi^{\text{tot}}_{\text{sp}}(\omega)$, which is defined in the Sec.~\ref{maintext:results}, as a function of $U$ in Fig.~\ref{Fig:apdx:6bathChi}. From Fig.~\ref{Fig:apdx:6bathChi} (a), it is evident that the low-energy local spin excitations which are responsible for the Kondo resonance persist in both metallic and the OSM phases in the case of full Hund's coupling. On the contrary, although the low-energy local spin excitations appear in metallic phase, those are absent once the system enters the OSM phase in the case of Ising Hund's coupling. Furthermore, the low-energy dynamical charge fluctuations are substantially suppressed (not shown) in the OSM phase for both cases. Hence, the low-energy physics of the two-orbital system in the OSM phase is dominated by low-energy spin excitations. Thereby, it is concluded that the distinct behavior of low-energy local spin excitations is responsible for the remarkable difference in the OSM phase between full and Ising Hund's coupling.

\begin{figure}[htbp]
\includegraphics[width=0.48\textwidth]{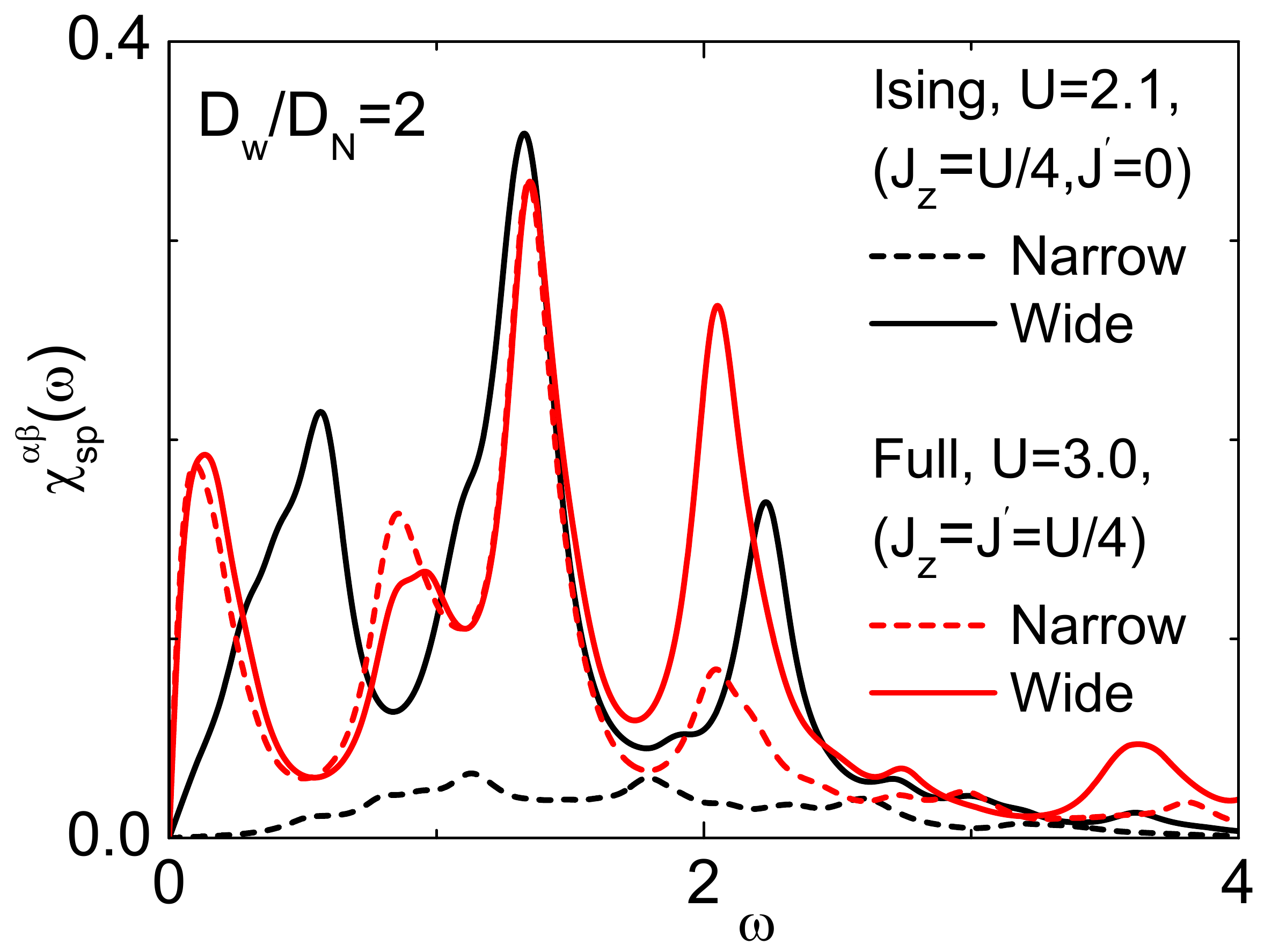}
\caption{(Color online) The orbitally resolved dynamical spin susceptibility $\chi_{sp}^{\alpha \beta}(\omega)$ in the OSM phases for both full (red) and Ising (black) Hund's coupling. Here, a Lorentz broadening factor of $0.1$ is used.}
\label{Fig:apdx:6bathspinSUS}
\end{figure}

In Fig.~\ref{Fig:apdx:6bathspinSUS}, we present the orbitally resolved dynamical spin susceptibility $\chi_{sp}^{\alpha \beta}(\omega)$ as defined in the Sec.~\ref{maintext:results} in the OSM phases for both full and Ising Hund's coupling. We find that in the Fermi-liquid OSM phase, the spin excitations are both gapless in wide and narrow bands even though narrow band is insulating, while in the non-Fermi-liquid OSM phase, spin excitations are gapped in narrow band but remain gapless in wide band. These indicate that although charge degrees of freedom between orbitals are decoupled in both OSM phases~\cite{MediciPRB2011}, the spin degrees of freedom behave distinctly, dependent on whether the coupling between orbitals is full or of Ising type. If full, the spins of different orbitals are strongly coupled with each other. Otherwise, the spins are decoupled between orbitals and the spin in narrow band acts as an effective magnetic field relative to that in wide band. Therefore, the widely accepted scenario for the OSM phase~\cite{MediciPRB2011} should be revised.

\subsection{ground state energy\label{subsect:energy}}
\begin{figure}[htbp]
\includegraphics[width=0.48\textwidth]{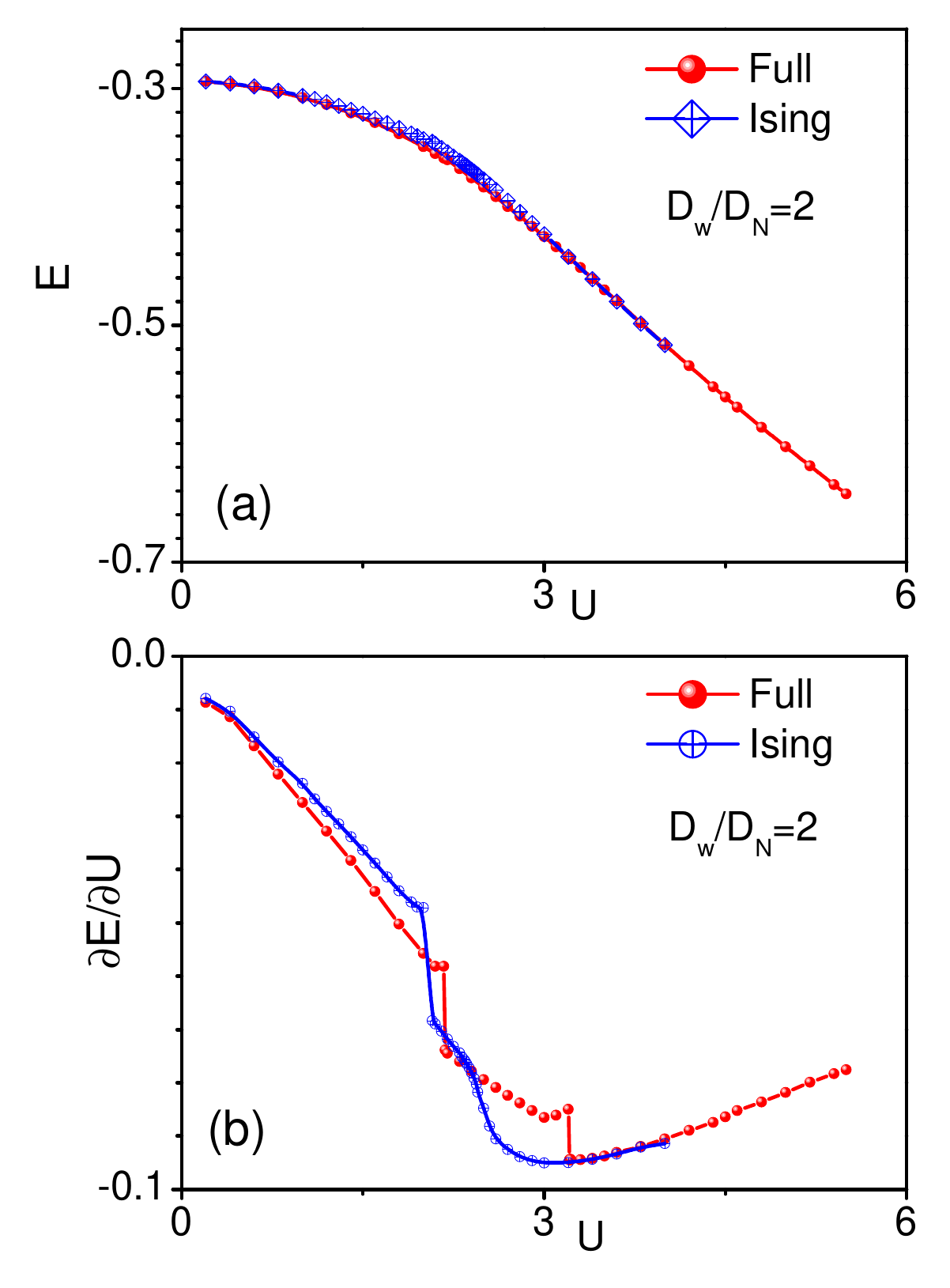}
\caption{(color online) The groundstate energy (a) and the first derivative of the groundstate energy (b) with respect to $U$ for full and Ising Hund's coupling as a function of interaction $U$.}
\label{Fig:apdx:GroundEg}
\end{figure}

Fig.~\ref{Fig:apdx:GroundEg} shows groundstate energy and its first derivative with respect to $U$ for full and Ising Hund's coupling as a function of $U$. It is found in Fig.~\ref{Fig:apdx:GroundEg} (b) that the first derivative exhibits two discontinuities for full Hund's coupling as $U$ is increased, while it displays only one discontinuity for Ising Hund's coupling, indicating two first-order phase transitions happening for full Hund's coupling and one for Ising limit. Further doing second derivative of groundstate energy with respect to $U$ for Ising Hund's coupling, we find one more discontinuity at larger $U$, indicating the occurrence of a second-order phase transition. All the critical values and the nature of the phase transitions are consistent with those derived from entanglement entropy.

However, as has been widely noticed that the entanglement entropy which measures the strength of quantum fluctuations is sensitive to the presence of quantum phase transitions~\cite{AmicoRMP2008}, we find that it is also the case for the OSM transitions. In the OSM phase, orbital fluctuations are suppressed, leading to an abrupt reduction in the entanglement entropy at the metal-to-OSM phase transition, which gives clearer indication for a strong first-order phase transition than the groundstate energy does. In the Mott phases, the charge fluctuations in both orbitals are fully suppressed, resulting in a further decreasing of the entanglement entropy. For full Hund's coupling, while groundstate energy varies smoothly at the OSM-to-Mott phase transition, the entanglement entropy shows a clear slope change, indicating an apparent discontinuity in the first derivative of the entanglement entropy with respect to $U$ and therefore a first-order phase transition. To our knowledge, we for the first time report the relation between derivatives of the entanglement entropy and the order of the phase transition.

Finally, we should point out another reason why we use entanglement entropy to determine the quantum OSM transitions, rather than the other physical quantities, such as the double occupancy in each orbital or interorbital correlation. This is due to the fact that the entanglement entropy is the only quantity which captures simultaneously the interorbital and intra-orbital charge fluctuations, as well as the spin fluctuations.

\section{Finite size effect\label{sect:three}}
\subsection{comparison of entanglement entropies obtained from 12 bath sites and 8 bath sites\label{subsect:comparisonEE}}
\begin{figure}[htbp]
\includegraphics[width=0.48\textwidth]{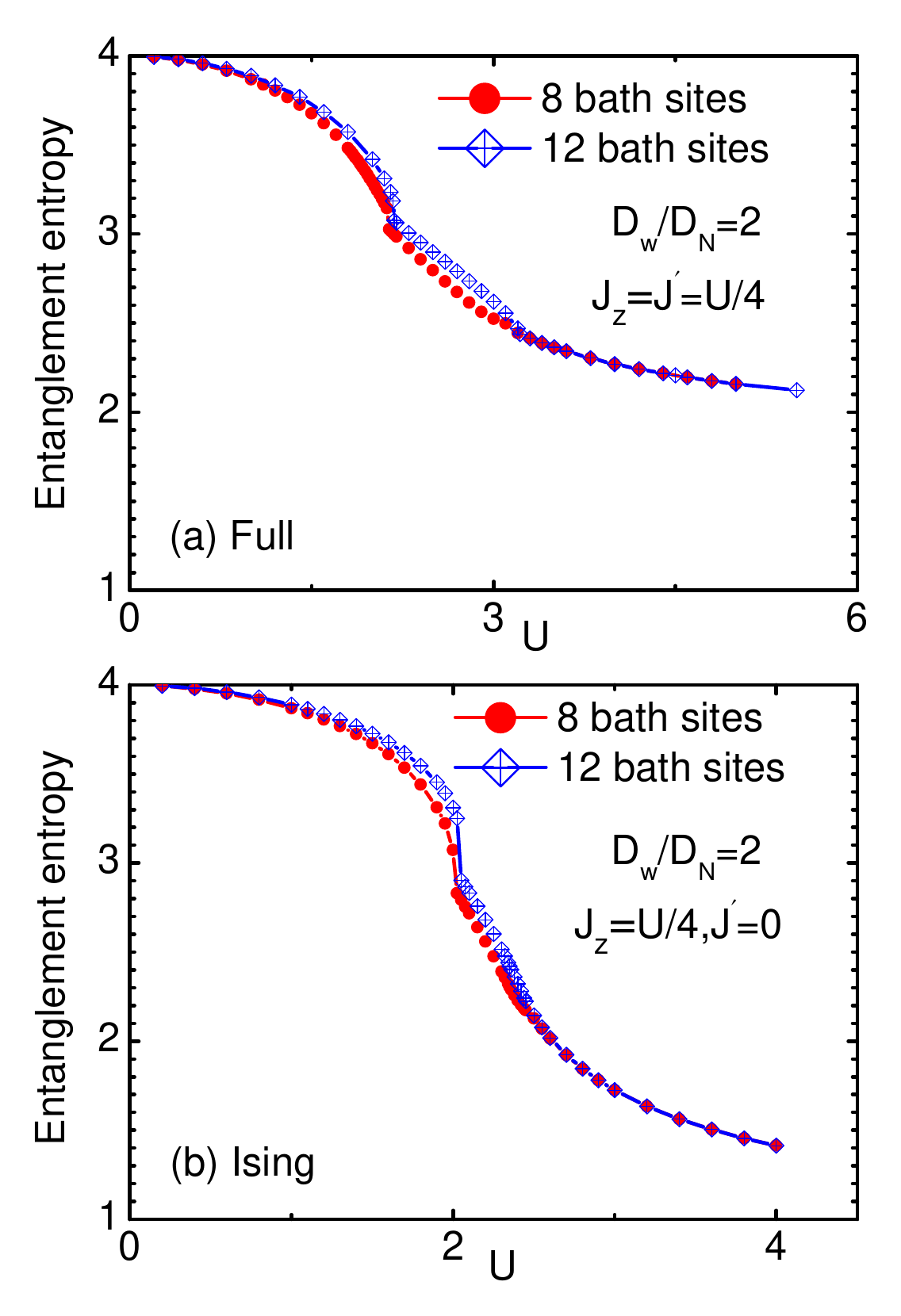}
\caption{(color online) Comparison of the entanglement entropies as a function of $U$ for full and Ising Hund's coupling calculated in the framework of DMFT with ED solver where 6 bath sites and 4 bath sites coupled to each impurity, namely 12 bath sites and 8 bath sites in total, respectively, are taken into account.}
\label{Fig:apdx:entropy}
\end{figure}

In Fig.~\ref{Fig:apdx:entropy}, we show a comparison of the entanglement entropies calculated in the framework of DMFT with 6 bath sites coupled to each impurity and 4 bath sites coupled to each impurity, namely 12 bath sites and 8 bath sites in total, respectively, in the ED solver for both full and Ising Hund's coupling. Only small deviations are found between 6 bath sites per impurity and 4 bath sites per impurity in metallic and the OSM phases, which does not affect the critical values and the natures of phase transitions. Moreover, as we show in Appendix~\ref{subsect:4bath} where DOS, imaginary part of self-energy, and dynamical spin susceptibility are calculated within DMFT by using 4 bath sites per impurity in ED solver, the natures of metallic, OSM, and insulating phases are not affected by the number of bath sites as well. Please note, we will give the reason why there is a small deviation in next section.

\subsection{effect of bath discretization\label{subsect:bath}}
\begin{figure}[htbp]
\includegraphics[width=0.48\textwidth]{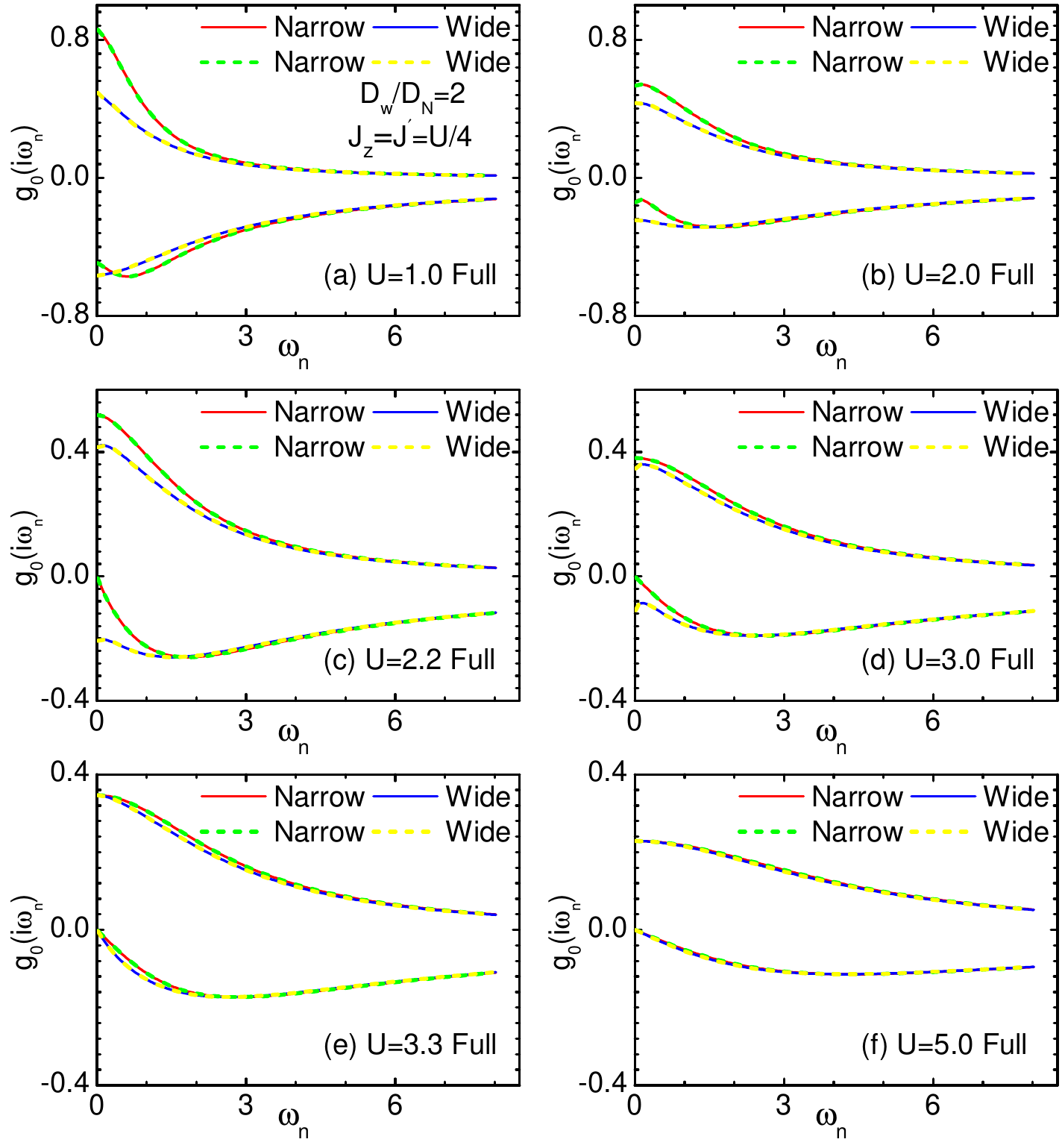}
\caption{(color online) The comparison between $\hat{g}_{0\gamma}(i\omega_n)$ and $g_{0\gamma}(i\omega_n)$ calculated by 12 bath sites (6 bath sites per impurity) for full Hund's coupling. Solid lines denote $g_{0\gamma}(i\omega_n)$ and dashed lines denote $\hat{g}_{0\gamma}(i\omega_n)$. The dashed and solid lines are completely overlapped, indicating negligible effect of bath discretization.}
\label{Fig:apdx:weissfdHund}
\end{figure}
\begin{figure}[htbp]
\includegraphics[width=0.48\textwidth]{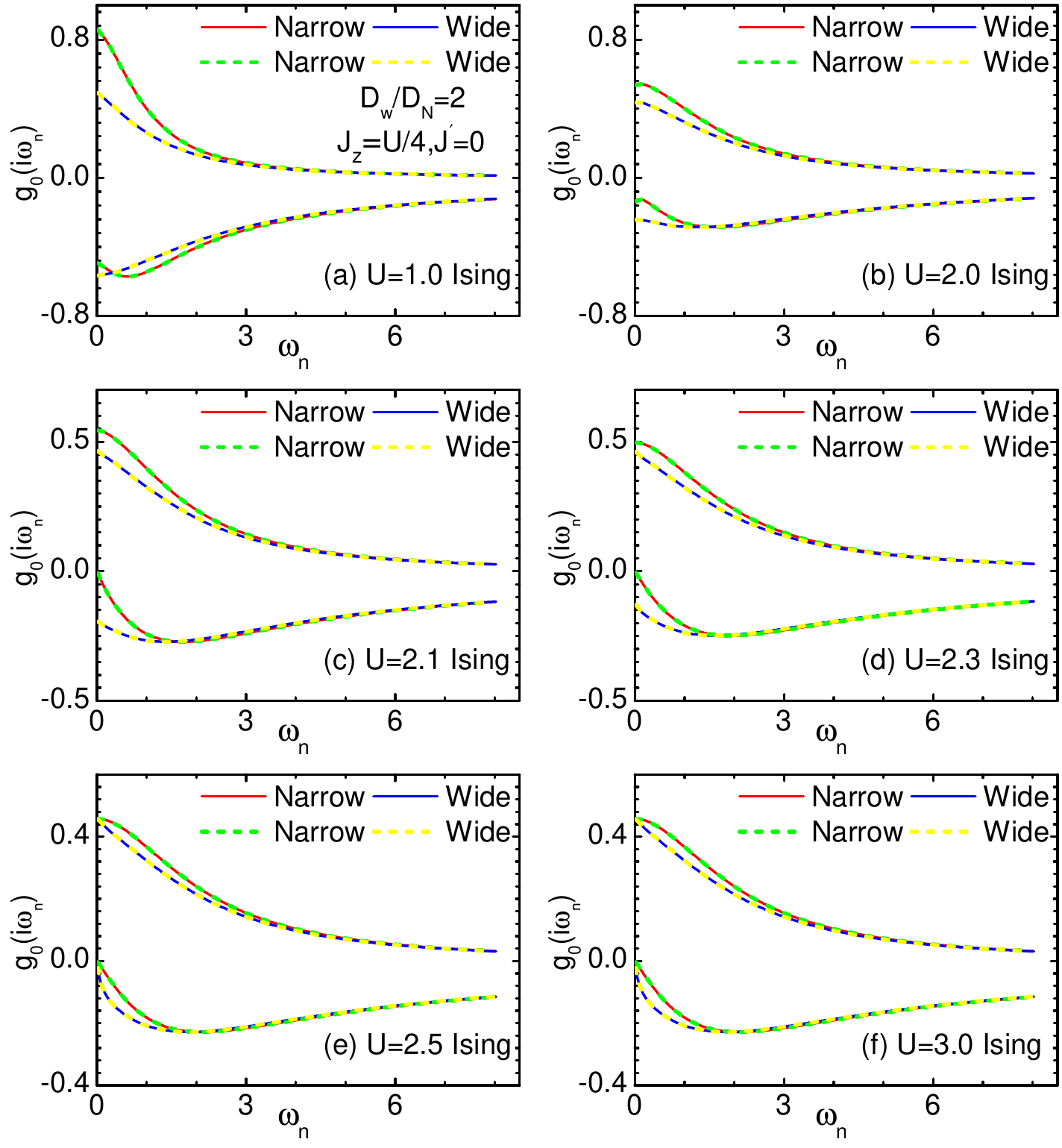}
\caption{(color online) The comparison between $\hat{g}_{0\gamma}(i\omega_n)$ and $g_{0\gamma}(i\omega_n)$ computed by 12 bath sites, namely 6 bath sites per impurity, for Ising Hund's coupling. Solid lines denote $g_{0\gamma}(i\omega_n)$ and dashed lines denote $\hat{g}_{0\gamma}(i\omega_n)$. The dashed and solid lines are completely overlapped, indicating negligible effect of bath discretization.}
\label{Fig:apdx:weissfdIsing}
\end{figure}

A crucial step in the framework of DMFT combined with ED solver is the bath discretization. We need to search a set of optimized bath parameters which can minimize the difference between the Weiss field of the two-orbital Hubbard model $g_{0\gamma}(i\omega_n)$ and the noninteracting impurity Green's function of the discretized two-orbital Anderson impurity model $\hat{g}_{0\gamma}(i\omega_n)$. As shown in Fig.~\ref{Fig:apdx:weissfdHund}, we find that $\hat{g}_{0\gamma}(i\omega_n)$ derived from the discretized two-orbital Anderson impurity model with 12 optimized bath sites, namely 6 bath sites per impurity, can reproduce $g_{0\gamma}(i\omega_n)$ quantitatively well in all three phases, including metallic phase (Fig.~\ref{Fig:apdx:weissfdHund} (a) and (b)), the OSM phase (Fig.~\ref{Fig:apdx:weissfdHund} (c) and (d)), and Mott insulating phase (Fig.~\ref{Fig:apdx:weissfdHund} (e) and (f)) in the case of full Hund's coupling. Likewise, $g_{0\gamma}(i\omega_n)$ can also be reproduced by $\hat{g}_{0\gamma}(i\omega_n)$ in the case of Ising Hund's coupling as depicted in Fig.~\ref{Fig:apdx:weissfdIsing}. Our results suggest that the effect of bath discretization is negligible when 6 bath sites per impurity is used. And 6 bath sites per impurity is large enough to give reliable results quantitatively.

In addition, we should point out that the small deviations found in Fig.~\ref{Fig:apdx:entropy} is due to small deviations between $g_{0\gamma}(i\omega_n)$ and $\hat{g}_{0\gamma}(i\omega_n)$ when 4 bath sites per impurity is used. Combining the results shown in Appendix~\ref{subsect:comparisonEE} and following  Appendix~\ref{subsect:4bath}, we can conclude that even 4 bath sites per impurity is large enough to give reliable results qualitatively.

In fact, in Ref.~\cite{GeorgesRMP1996} it was emphasized that the ED solver in the framework of DMFT does not deal with a finite-size lattice for the original lattice model. The discretization concerns only the effective conduction bath in the impurity-model formulation. Therefore, as long as one can fit the Weiss field of lattice model by the non-interacting Green's function of the Anderson impurity model with optimized discretized bath parameters, the finite-size effect, or precisely speaking, the effect of bath discretization, can be ignored.

\subsection{results with 4 bath sites coupled to each orbital\label{subsect:4bath}}
In the Sec.~\ref{maintext:results}, all the calculations in the framework of DMFT combining with ED are carried out with totally 12 bath sites. In order to clarify whether the given results depend on the number of bath sites, we had performed the calculations with totally 8 bath sites, namely each orbital coupled to 4 bath sites, at bandwidth ratio of $D_W:D_N=2:1$. For comparison to the results of 12 bath sites, the orbital resolved DOS, $Im\Sigma(i\omega_n)$, static and dynamical spin susceptibility $\chi^{\text{tot}}_{sp}(\omega)$ are also calculated. All the results from totally 8 bath sites agree qualitatively well with those from totally 12 bath sites for all $U$.
\subsubsection{density of states\label{4bathDOS}}
\begin{figure}[htbp]
\includegraphics[width=0.48\textwidth]{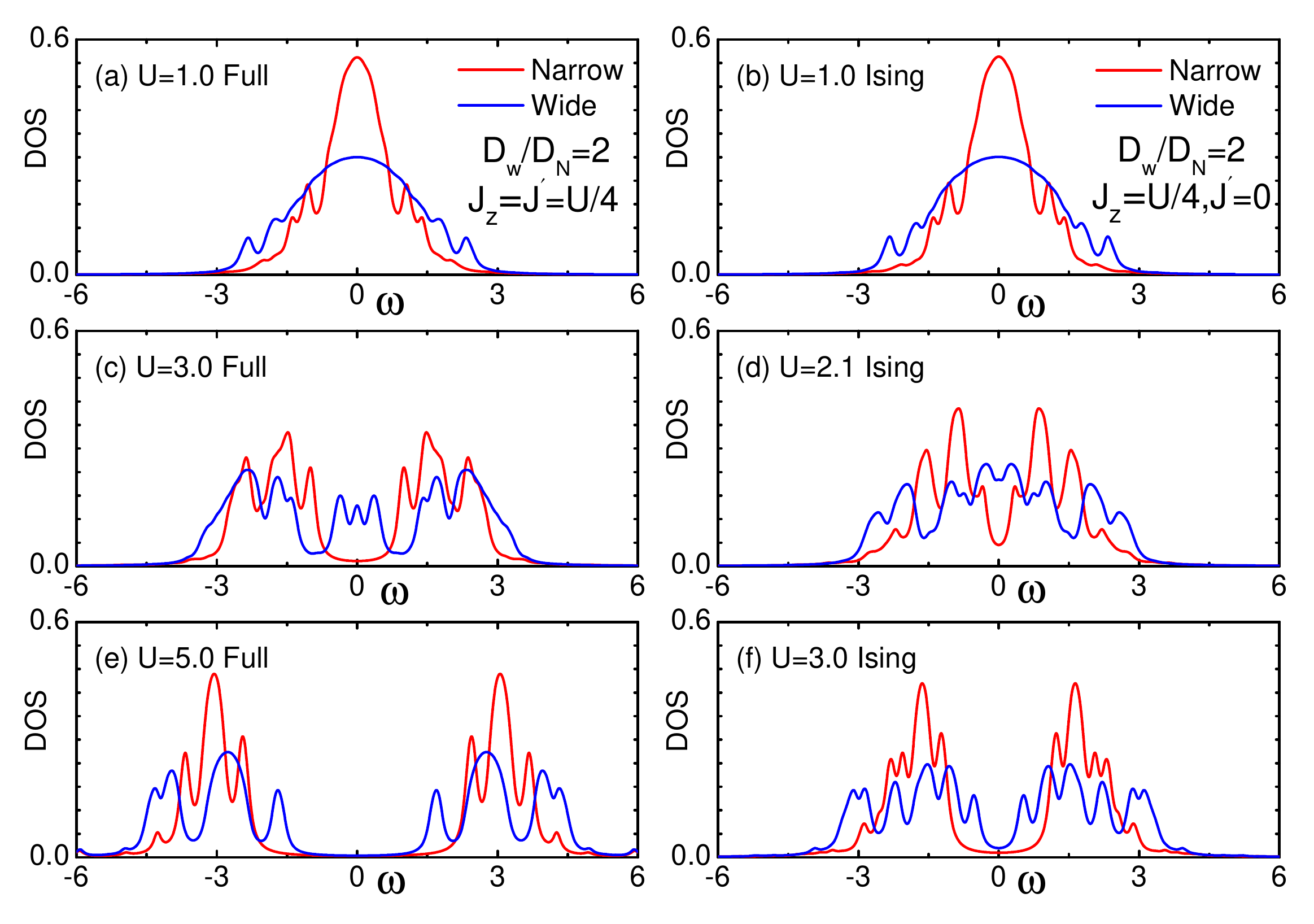}
\caption{(Color online) The orbital-resolved density of states in metallic phase (a), (b), the OSM phase (c), (d), and Mott insulating phase (e), (f). Here, a Lorentz broadening factor of $0.1$ is used.}
\label{Fig:apdx:4bathDOS}
\end{figure}

Fig.~\ref{Fig:apdx:4bathDOS} displays the DOS produced by ED solver with totally 8 bath sites. At small $U$ as shown in Fig.~\ref{Fig:apdx:4bathDOS} (a) and (b), both bands have finite DOS at the Fermi level, indicating metallic state in two bands for both types of Hund's coupling. When the onsite Coulomb interaction is strong enough, the system enters insulating phase as shown in Fig.~\ref{Fig:apdx:4bathDOS} (e) and (f). The OSM phase is also observed with totally 8 bath sites at intermediate values of $U$ in both full and Ising Hund's coupling, e.g. seen in Fig.~\ref{Fig:apdx:4bathDOS} (c) and (d). In addition, a central peak at the Fermi level in wide band in the OSM phase is clearly present in the case of full Hund's coupling as shown in Fig.~\ref{Fig:apdx:4bathDOS} (c), while a dip at the Fermi level appears in the case of Ising Hund's coupling as seen in Fig~\ref{Fig:apdx:4bathDOS} (d). The results are consistent with those obtained from totally 12 bath sites.

\begin{figure}[htbp]
\includegraphics[width=0.48\textwidth]{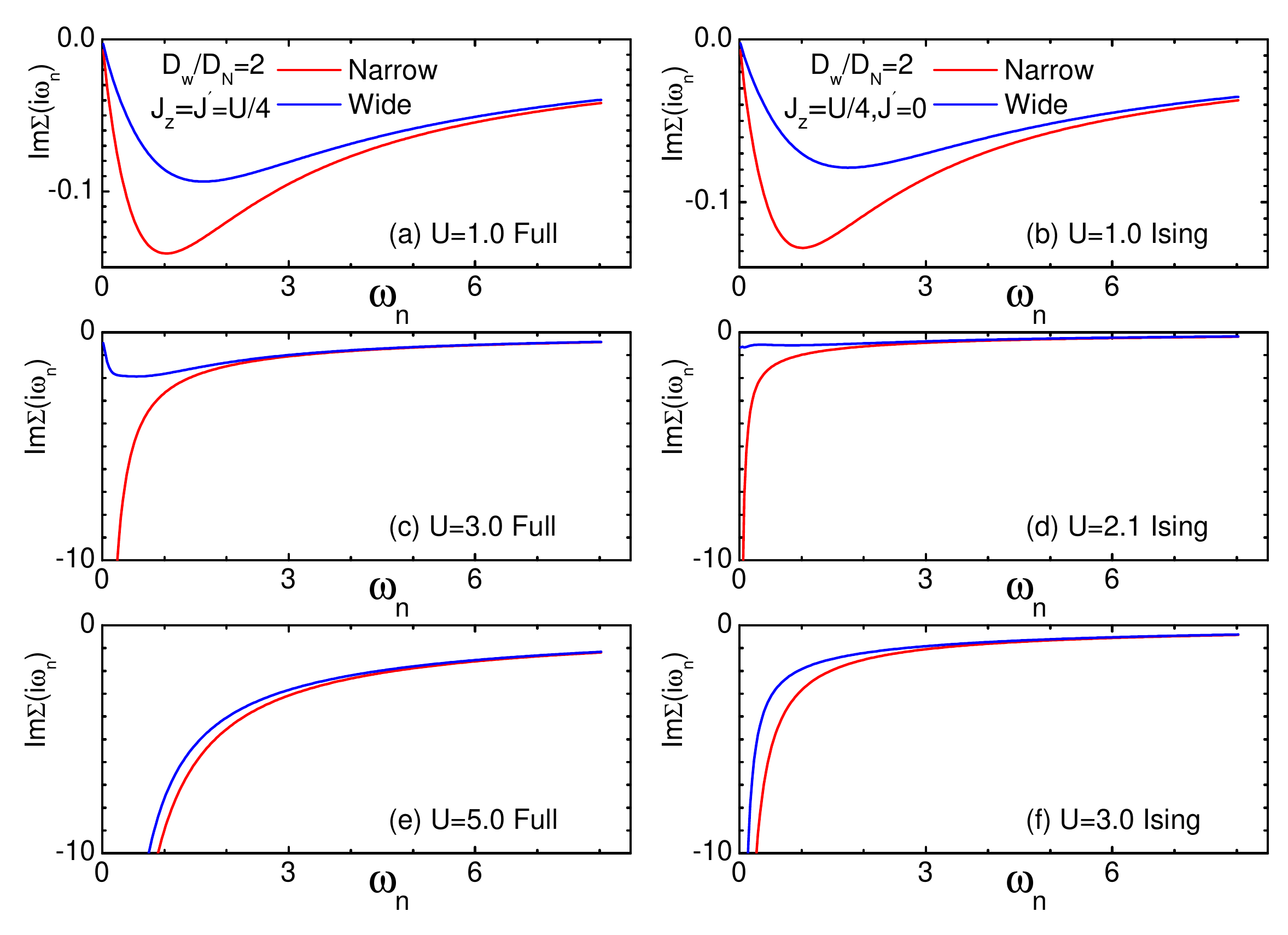}
\caption{(Color online) The imaginary part of self-energy in Matsubara frequency $Im\Sigma(i\omega_n)$ in metallic phase (a), (b), the OSM phase (c), (d), and Mott insulating phase (e), (f).}
\label{Fig:apdx:4bathSelfEg}
\end{figure}
\begin{figure}[htbp]
\includegraphics[width=0.48\textwidth]{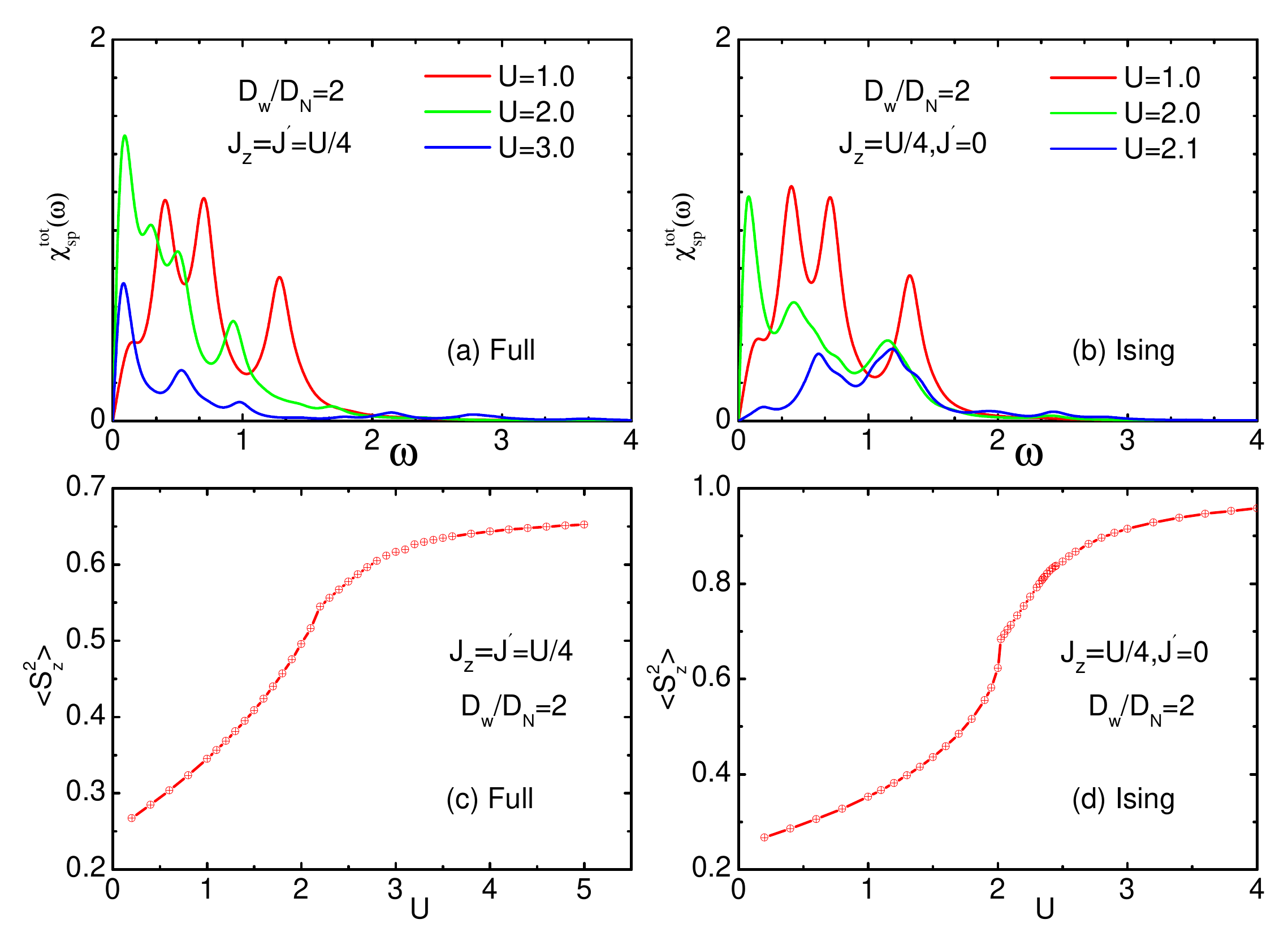}
\caption{(Color online) Dynamical spin susceptibility for full (a) and Ising (b) Hund's coupling, and static spin susceptibility for full (c) and Ising (d) Hund's coupling. In (a) and (b), a Lorentz broadening factor of $0.1$ is used.}
\label{Fig:apdx:4bathChi}
\end{figure}
\subsubsection{imaginary part of self-energy in Matsubara frequency\label{4bathselfenergy}}
The imaginary part of self-energy in Matsubara frequency $Im\Sigma(i\omega_n)$ is described in Fig.~\ref{Fig:apdx:4bathSelfEg} for both types of Hund's coupling. $Im\Sigma(i\omega_n)$ of both narrow and wide band approach zero as Matsubara frequency goes to zero in metallic phase for both types of Hund's coupling (Fig.~\ref{Fig:apdx:4bathSelfEg} (a) and (b)), implying both are the Fermi liquid, while all
quantities diverge proximity to zero frequency in insulating phase (Fig.~\ref{Fig:apdx:4bathSelfEg} (e) and (f)), which is the typical behavior of Mott insulator. In the OSM phase for each type of Hund's coupling, only the imaginary part of self-energy in Matsubara frequency of narrow band diverge
(Fig.~\ref{Fig:apdx:4bathSelfEg} (c) and (d)). The difference between full and Ising Hund's coupling is that the imaginary part of self-energy in Matsubara frequency of wide band in the OSM phase goes to zero as the
Matsubara frequency goes to zero in the case of full Hund's coupling, while it approaches a finite value in the case of Ising Hund's couplng. The results again agree well with those given by 12 bath sites.

\subsubsection{dynamical spin susceptibility\label{4bathdynamical}}
As shown in Fig.~\ref{Fig:apdx:4bathChi}, for full Hund's coupling, the low-energy local spin excitations persist in the OSM phase, which is in sharp contrast to the absence of low-energy local spin excitations in the OSM phase for Ising Hund's coupling. In addition, the static spin fluctuations approach $\frac{2}{3}$ indicating the formation of local spin triplet states in the case of full Hund's coupling, and those approach $1$ suggesting the formation of local spin doublet states in the case of Ising Hund's coupling, in the strong coupling limit. The results are the same as the results from 12 bath sites.

\end{document}